\documentclass[]{jfm}

\usepackage{amsmath,bm}

\usepackage{graphicx}
\usepackage{placeins}
\usepackage{subfig}
\usepackage[percent]{overpic} 
\usepackage{mathtools}
\usepackage{newtxtext}
\usepackage{newtxmath}
\usepackage{natbib}
\usepackage{hyperref}
\usepackage{marginnote}
\hypersetup{
    colorlinks = true,
    urlcolor   = blue,
    citecolor  = black,
}

\newcommand{\RomanNumeralCaps}[1]
\linenumbers

\usepackage{subfiles}

\newcommand{\bA}{\bm{A}}
\newcommand{\bB}{\bm{B}}
\newcommand{\bS}{\bm{S}}
\newcommand{\bu}{\bm{u}}

\newcommand{\tD}{{{\rm D}}}
\newcommand{\ez}{{\bm{e}}_z}

\title{Impact of alignments between fluctuating and mean density gradients on the scale-dependent energetics of stably stratified turbulence}

\author{Soumak Bhattacharjee\aff{1},
  Stephen M. de Bruyn Kops\aff{2}
 \and Andrew D. Bragg \aff{1}
  \corresp{\email{andrew.bragg@duke.edu}}}

\affiliation{\aff{1} Department of Civil and Environmental Engineering, Duke University, Durham, NC 27708, USA
\aff{2} Department of Mechanical and Industrial Engineering, University of Massachusetts Amherst,
Amherst, MA 01003, USA}

\begin{document}
\maketitle

\begin{abstract}

Non-trivial alignments between vorticity and the strain-rate tensor play an important role in the evolution of velocity gradients and the energy cascade in isotropic turbulence. Here we explore how alignments between the fluctuating and mean density gradients impact the mechanisms governing the turbulent kinetic energy (TKE) and available potential energy (APE) across scales in stably stratified turbulence. This is motivated by analytical results that demonstrate a connection between them, and is conducted using direct numerical simulations (DNS) of statistically stationary, stably stratified turbulence for $Pr = 1, 7, 50$ in the strongly stratified regime. After demonstrating how the gradient field alignments depend on scale and $Pr$, we show that the alignments are intimately connected to the reversal of the buoyancy flux at small-scales, and that regions of strong alignment and misalignment correspond to regions where the horizontal TKE inter-scale flux becomes weak. The same is also true of the APE flux, except that at larger scales, regions of strong alignment are associated with an upscale APE flux. The TKE and APE dissipation rates, and the mixing coefficient also show a strong dependence on the alignment, especially for $Pr=1$. Finally, we explore the connection between the local alignment and stability of the flow, and we find a non-trivial relationship, with regions of strong alignment surprisingly occurring most often in stable regions. This demonstrates that the dynamical significance of the alignments on the flow energetics cannot be understood through a simple connection between the local alignments and local stability of the flow.

\end{abstract}

\begin{keywords}
\end{keywords}

{\bf MSC Codes }  {\it(Optional)} Please enter your MSC Codes here

\section{Introduction}\label{sec:Introduction}


Turbulence in geophysical flows is often strongly influenced by buoyancy forces due to temperature and/or salinity differences which generate gradients in the fluid density field. In the case where the density is stably stratified, buoyancy forces permit internal waves, and inhibit vertical fluid motion, making turbulence highly anisotropic and spatio-temporally intermittent. Understanding and modelling stratified turbulence is a major challenge with many open questions; see  \cite{Harindra1991,Staquet1996,RileyLindborg_2012,Caulfield_2020,Caulfield_2021} for overviews. 

In neutrally buoyant flows such as homogeneous, isotropic turbulence (HIT), preferential alignments among various flow quantities are known to be of fundamental importance to the basic dynamics and energetics of the flow. It is well known that the vorticity preferentially aligns with the intermediate strain-rate eigenvector \citep{ashurst_alignment_1987,tsinober_book,meneveau_lagrangian_2011}, which is surprising since standard arguments suggest that it would be aligned with the extensional strain-rate eigenvector \citep{Xu2011}. This non-trivial alignment with the intermediate eigenvector has significant consequences for turbulence, as it implies that enstrophy production is not as strong as it could be. The same basic alignment property also holds for the filtered vorticity and strain-rate fields, and given the contribution of the filtered enstrophy production to the kinetic energy cascade \citep{carbone_bragg_2020,Johnson_2021,Johnson_2020}, this means that the energy cascade is less efficient than it might otherwise be \citep{Ballouz_Ouellette_2018}. Recent research has also discovered a self-attenuation mechanism according to which the spatially local contribution to the enstrophy production term acts to prevent singular growth of enstrophy due to a local preferential alignment of the vorticity with the velocity vector \citep{Buaria_2020}. This is consistent with the finding that when the local flow helicity is high, the interscale energy transfer is reduced \citep{jacobitz_influence_2011}. For passive scalars in turbulence, preferential alignments and flow geometry also play an important role in determining scalar transport across flow scales, scalar-gradient amplification, and the eventual diffusion at small scales.
The scalar gradient preferentially aligns with the compressive strain-rate eigenvector \citep{ashurst_alignment_1987}. This is required in order for the fluctuating scalar gradients to be sustained in the flow against their dissipation-rate, and the alignment is also connected to the direction and efficiency of the passive scalar energy (or ``scalar variance'') cascade \citep{Zhang2022}. This preferential alignment also enhances the scalar iso-surface area through stretching along the extensional eigendirections, with strong consequences for the scalar mixing \citep{Shete_2019}. Moreover, when the passive scalar is driven by a mean scalar gradient, the fluctuating scalar field develops ramp-cliff structures \citep{Warhaft2000}, which are also associated with a preferential misalignment of the fluctuating density gradient with the mean density gradient \citep{BraggdBKops_2024}.
\cite{Buaria_2021,Shete2022} showed that as the scalar Prandtl number ($Pr\equiv \text{momentum diffusivity}/\text{scalar diffusivity}$) increases, the ramp-cliff structures weaken, and they predict that they vanish in the limit $Pr\to\infty$.

In the context of stratified flows, alignments between various flow quantities are also known to be of fundamental importance. For example, a classical result from geophysical fluid dynamics states that the potential vorticity is conserved along a fluid particle trajectory in an inviscid flow, where the potential vorticity is the inner product of the total vorticity and the density gradient \citep{davidson_turbulence_2013}. \cite{Watanabe_2016} found that stratified turbulent flows could be effectively partitioned into turbulent and non-turbulent regions based on the local potential enstrophy (which is the square of the potential vorticity) together with the enstrophy. \cite{Portwood2016} showed that a similar partitioning of the flow can be achieved based only on information on the vertical density gradient (i.e. the inner product of the density gradient with the vertical unit vector), since only regions with active turbulence are expected to generate locally unstable regions where the density gradient is positive.  

Using a stratified inclined duct experiment, \cite{Jiang_2023} recently illustrated how the alignment of the fluctuating density gradient relative to (a) the eigenvectors of the pseudo-dissipation tensor, and (b) the rortex and shear components of the vorticity field (which we denote by $\bm{\omega_r}$ and $\bm{\omega_s}$, respectively), dictates the efficiency of diapycnal mixing in sheared, stratified turbulence. The authors noted that preferential alignment of the fluctuating density gradient in the direction $\pm \bm{\omega_r}\times\bm{\omega_s}$ as well as the direction of maximal dissipation not only makes the isopycnals more susceptible to distortions through overturning by vortices, but also more exposed to stretching and diffusion, thereby increasing isosurface area. Both of these effects result in enhanced turbulent mixing.
\cite{Jiang_2025} also underscored the importance of stratification at sub-Ozmidov scales of high-Prandtl number flows, through the action of baroclinic torque on `rortex' structures (which are an order of magnitude smaller than the Ozmidov scale).

\cite{BraggdBKops_2024} established how the ramp-cliff structures in the density field affects the evolution of the mean turbulent kinetic energy (TKE) dissipation rate $\langle\epsilon\rangle$ and the mean available potential energy (APE) dissipation rate $\langle\chi\rangle$ in stably-stratified turbulence. In particular, \cite{BraggdBKops_2024} demonstrated that the preferential misalignment of the fluctuating density gradient with respect to the mean density gradient leads to a decrease in $\langle\epsilon\rangle$ and a simultaneous increase in $\langle\chi\rangle$. Using perturbation theory, they then showed how this explains previous observations that the mixing coefficient $\Gamma\equiv \langle\chi\rangle/\langle\epsilon\rangle$ decreases significantly with increasing Prandtl number $Pr$ in stratified turbulence \citep{Okino2017,Legaspi_2020,RileyCdBKops_2023}. \cite{Bhattacharjee_2026a} recently showed that the misalignment of the fluctuating filtered density gradient and mean density gradient is also responsible for the reversal of the scale-dependent mean buoyancy flux at small scales in stratified turbulence, which leads to the conversion of APE back into TKE at sub-Ozmidov scales, an effect that is important when $Pr\gg 1$.

The goal of this paper is to advance the analysis of \cite{Bhattacharjee_2026a} by considering how all of the terms in the scale-dependent mean TKE and APE equations depend on the local alignment between the fluctuating filtered density gradient and mean density gradient. As part of this analysis we also explore various alignments involving the fluctuating filtered density gradient, the filtered strain-rate eigenvectors, and filtered vorticity, since these alignments impact the behaviour of various terms in the TKE and APE equations. We also explore how the local alignment between the fluctuating filtered density gradient and mean density gradient is related to the notion of stability and instability at different scales in the flow, providing a connection between our geometric analysis, and notions of local stability that are more common in studies of stratified turbulence.

The rest of the paper is organised as follows: In \S 2 we present the theoretical analysis of the problem, and
in \S 3 we summarise the direct numerical simulations. In \S 4, we present results on the terms in the scale-dependent mean TKE and APE equations, first the unconditioned mean values, and then with the mean values conditioned on the local alignment between the fluctuating filtered density gradient and mean density gradient. Finally, in \S  5 we summarise the findings and discuss important future work.

\section{Theory}\label{sec:Theory}


\subsection{Governing scale-dependent equations}

We consider turbulence in a stably stratified Boussinesq fluid with a linear equation of state, where the total rescaled-density is
\begin{equation} \label{eq:density_profile}
    \Phi=\phi_r - N z + \phi,\; N^2 \equiv -g \zeta/\rho_r, \text{ with } \zeta < 0,
\end{equation}
where $\phi_r$ is the reference value contribution (associated with the constant reference density $\rho_r$), $-N\ez$ is the constant mean gradient contribution (associated with the constant mean density gradient $\zeta \ez$), $N$ is the Br\"unt-V\"ais\"al\"a frequency, $-g\ez$ is the gravitational acceleration, and $\phi$ is the fluctuating contribution (associated with the fluctuating density $\rho'$). Each of the contributions to $\Phi$ is associated with a different contribution to the total density $\rho=\rho_r-\zeta z+\rho'$ that has been rescaled so that it has dimensions of a velocity (which is often done to simplify the form of the governing equations).

To investigate the properties of the turbulence across scales, we define the filtering operation on an arbitrary field $ a(\bm{x},t)$ as
\begin{equation}\label{eq:def_filtering}
    \tilde{a}(\bm{x},t) \equiv \int_{\Omega} a(\bm{y},t)\mathcal{G}_{\ell}(\bm{x}-\bm{y})\,d\bm{y}.
\end{equation}
Here, $\mathcal{G}_{\ell}$ is a filter-kernel with characteristic scale $\ell$ \citep{VremanGeurtsKuerten_1994}, and $\Omega$ is the domain of the flow. 
The filtered field $\tilde{a}$ can be thought to contain the field information at scales $\gtrsim \ell$, whereas the sub-filter-scale (SFS) information is captured in $a - \tilde{a}$, see \S \ref{sec:Numerical} for details on the filter kernel choice. \\

Applying the filtering operator to the Boussinesq-Navier-Stokes equations we obtain the equations governing the dynamics of the filtered fields $\tilde{\bu}$ and $\tilde{\phi}$
\begin{align}
\tD_t \tilde{\bu} &=-\left(1 / \rho_r\right) \nabla \tilde{p}-\nabla \bm{\cdot} \bm{\tau}+2 \nu \bm{\nabla} \bm{\cdot} \tilde{\bm{S}}-N \tilde{\phi} \ez+\tilde{\bm{F}}, \label{eq:filteredNSE_boussinesqA}\\
\tD_t \tilde{\phi} &=-\nabla \bm{\cdot} \bm{\Sigma}+\kappa \nabla^2 \tilde{\phi}+N \tilde{u}_z. \label{eq:filteredNSE_boussinesqB}
\end{align}
Here, $\tD_t \equiv \partial_t + \tilde{\bu} \,\bm{\cdot}\bm\nabla$ denotes the filtered material derivative operator, $\bu$ is the velocity field with horizontal and vertical components $\bu_h$ and $u_z\equiv \bu\boldsymbol{\cdot}\ez$, respectively, $p$ is the pressure field, $\nu$ is the kinematic viscosity, $ \bm{S}\equiv(\nabla \bu + \nabla \bu^\top)/2$ denotes the strain-rate field, and $\kappa= \nu/Pr$ is the scalar diffusion coefficient.
Further, $\bm{F}$ is a forcing term which may be used to generate a statistically stationary state by compensating for the energy lost due to the dissipative mechanisms in the flow. 

The sub-filter stresses $\bm{\tau}$ and $\bm{\Sigma}$ are given by
\begin{equation*}
    \tau\equiv \widetilde{\bu \bu}-\tilde{\bu}\tilde{\bu}, \; \Sigma\equiv \widetilde{\bu \phi}-\tilde{\bu}\tilde{\phi}.
\end{equation*}
Note that for $\ell =0$, Eqns. \ref{eq:filteredNSE_boussinesqA},  \ref{eq:filteredNSE_boussinesqB} correspond to the unfiltered Boussinesq-Navier-Stokes equations with
\begin{equation*}
    \tilde{\bu}_{\ell=0} = \bu,\; \tilde{\phi}_{\ell=0} = \phi,\; \bm{\tau}=0,\; \bm{\Sigma}=0.
\end{equation*}

\subsection{Parameters and characteristic scales}

One of the central challenges in characterising stratified flows is the large parameter space of dimensionless numbers that affect the flow dynamics.
In contrast to (neutrally buoyant) homogeneous, isotropic turbulence (HIT) which can be solely characterised by the Reynolds number $Re \equiv u' L/\nu$ ($u'$ and $L$ being the characteristic velocity and length scales, respectively),
stratified turbulence is additionally characterised by the Froude number $Fr=u'/NL_h$ ($u'$ and $L_h$ being the characteristic horizontal velocity and length scales, respectively),
and the Prandtl number $Pr=\nu/\kappa$ (or Schmidt number $Sc$). We do not distinguish between these within the context of the governing equations under consideration and the fluid may be considered to be stratified due to temperature or salinity gradients, for example). The Froude number indicates the strength of static stability, with buoyancy being unimportant for $Fr \gg 1$ (the passive scalar limit), and of leading order importance for $Fr \leq O(1)$.
The Prandtl number determines the scale-separation between the viscous scale for $\bm{u}$, given by the Kolmogorov length scale $\eta$ \citep{kolmogorov41c}, and the diffusion scale for $\phi$. We confine attention to $Pr\geq O(1)$ for which the latter scale is the Batchelor length scale $\eta_B \equiv \eta/Pr^{1/2}$ \citep{Batchelor1959}.
In the context of passive scalars, a viscous-convective range exists only if $Pr^{1/2} \gg 1$, such that for scales in the range $\eta \gg \ell \gg \eta_B$, the viscosity is important for the flow dynamics but the diffusivity is not.


Stratification plays a key role in suppressing turbulence at scales greater than the Ozmidov scale $l_{O}\equiv(\langle \epsilon \rangle/N^3)^{1/2}$, and becomes sub-leading at scales $\ell\ll l_{O}$.
A question of fundamental importance is whether stratification affects the smallest scales of the turbulent flow.
The activity parameter, $Gn \equiv \langle \epsilon \rangle /(\nu N^2) = (l_O/\eta)^{4/3}$, is an important dimensionless number quantifying how large the Ozmidov scale is compared to $\eta$.
$Gn \gg 1$ implies a large-separation of scales ($l_O \gg \eta$), and three-dimensional Kolmogorov turbulence is expected to emerge in the sub-Ozmidov inertial range $l_O \gg \ell \gg \eta$ \citep{RileyLindborg_2012}.
The parameter $Gn$ is closely related to the buoyancy Reynolds number $Re_b \equiv Re Fr^2$, which has traditionally been used to understand the effect of stratification on the small-scales \citep{BillantChomaz2001,RileydBK2003,RileyLindborg_2012}.
However, the analysis in \cite{BraggdBKops_2024} suggests that this standard view of stratified turbulence is not correct in the regime $Pr\gg1$, because in that regime the effect of buoyancy at the small-scales of the flow can be strong even if $Gn\gg1$.

\subsection{Equations governing the scale-dependent energetics}

The energy in the fluctuating fields at scales $\geq \ell$ is given by the filter-scale (FS) energies \citep{Germano_1992}
\begin{equation}
    E_h \equiv \frac{1}{2}\|\tilde{\bu}_h \|^2,\; E_z \equiv  \frac{1}{2}\tilde{u}_z^2,\; E_\phi \equiv  \frac{1}{2}\tilde{\phi}^2, \label{eq:fsE}
\end{equation}
where $E_h$, $E_z$ and $E_\phi$ are the horizontal and vertical components of the FS TKE, and the FS APE, respectively. (Strictly, this is not the APE defined in \citet{Winters_1995}, however, it is usually used to approximate it, with the difference between them being small for small density fluctuations, see
\citet{Howland_Taylor_Caulfield_2021}). Similarly, the energy associated with scales $< \ell$ is given by the sub-filter-scale (SFS) energies
\begin{equation}
    e_h \equiv  \frac{1}{2}\widetilde{\|\bu_h\|^2}-E_h,\; e_z \equiv  \frac{1}{2}\widetilde{u_z^2}-E_z,\; e_\phi \equiv  \frac{1}{2}\widetilde{\phi^2} - E_\phi,\label{eq:sfsE}
\end{equation}
where $e_h$, $e_z$ and $e_\phi$ are the horizontal and vertical components of the SFS TKE, and the SFS APE, respectively.

The evolution of the ensemble-averaged FS energies \eqref{eq:fsE}, assuming statistical homogeneity, is given by
\begin{align}
    \partial_t \langle E^{
    }_{h} \rangle &= \langle P^f_h \rangle 
    - \langle \Pi_{h}\rangle
    - \langle \Pi_{r,h}\rangle -\langle \tilde{\epsilon}_{h}\rangle -\langle\tilde{\epsilon}_{r,h}\rangle +\langle\mathcal{F}^f_{K}\rangle, \label{eq:FStke_budget_ss1}\\
    \partial_t \langle E^{
    }_{z} \rangle &= \langle P^f_z \rangle -\langle \Pi_{z}\rangle
    - \langle \Pi_{r,z}\rangle -\langle\tilde{\epsilon}_{z}\rangle -\langle\tilde{\epsilon}_{r,z}\rangle - \langle \mathcal{B}^f\rangle , \label{eq:FStke_budget_ss2}\\
    \partial_t \langle E^{
    }_\phi \rangle &= \langle \mathcal{B}^f\rangle -\langle \Pi^f_\phi\rangle -\langle\tilde{\chi}\rangle \label{eq:FStke_budget_ss3},
\end{align}
where, $P^f_h$ and $P^f_z$ are the pressure-redistribution terms
\begin{align}
    P^f_h &\equiv \tilde{p} \nabla_h \bm{\cdot} \tilde{\bu}_h /\rho_r,\; P^f_z \equiv \tilde{p} \nabla_z \tilde{u}_z /\rho_r,\;
    P^f_h + P^f_z = 0,
\end{align}
$\Pi_{h}$, $\Pi_{z}$ and $\Pi_\phi$ are the scale-to-scale horizontal TKE, vertical TKE, and APE fluxes
\begin{align}
    \Pi_h &\equiv - {\tau}_{hj}  \tilde{{S}}_{hj},\;\;\; \Pi_z\equiv - {\tau}_{zj} \tilde{{S}}_{zj},\;
    \Pi_K \equiv - \bm{\tau} \bm{:}  \tilde{\bm{S}}=\Pi_h + \Pi_h,\; \Pi_\phi = -\bm{\Sigma} \bm{\cdot} \tilde{\bB},
\end{align}
where $\tilde{\bB} \equiv \nabla \tilde{\phi}$. Note that we have defined $\Pi_h$ and $\Pi_z$ such that they only contain the strain-rate portion of the velocity gradient tensor, so that these terms can be physically interpreted as true scale-to-scale energy fluxes associated with a strain-rate and stress. The contributions involving the rotation-rate $\bm{R}\equiv(\nabla \bu - \nabla \bu^\top)/2$ are defined as
\begin{align}
    \Pi_{r,h} &\equiv - {\tau}_{hj}  \tilde{{R}}_{hj},\;\;\; \Pi_{r,z} \equiv - {\tau}_{zj}  \tilde{{R}}_{zj},\; \Pi_{r,h} + \Pi_{r,z} = 0.  
\end{align}
These terms appear with opposite signs in the filtered and sub-filter scale equations, meaning that these terms do transfer energy across scales. However, whereas $\Pi_h +\Pi_z$ is the total TKE flux which can be of arbitrary magnitude and sign, $\Pi_{r,h}+\Pi_{r,z}=0$ such that these terms act to redistribute energy among the scales. For example, if horizontal TKE is being transferred downscale due to $\Pi_{r,h}>0$, then we must have $\Pi_{r,z}<0$ corresponding to upscale transfer of vertical TKE. If this contribution is larger in magnitude than $\Pi_z$ then $\Pi_{r,z}+\Pi_z<0$, implying that nonlinearity leads to an overall upscale transfer of vertical TKE even if $\Pi_z>0$.

$\tilde{\epsilon}_{h}$ and $\tilde{\epsilon}_{z}$ are the FS horizontal and vertical TKE dissipation rates
\begin{align}
    \tilde{\epsilon}_{h} & \equiv 2 \nu \tilde{{S}}_{hj} \tilde{{S}}_{hj},
    \tilde{\epsilon}_{z} \equiv 2 \nu \tilde{{S}}_{zj} \tilde{{S}}_{zj},\; \;
    \tilde{\epsilon} \equiv  2 \nu \|\tilde{\bm{S}}\|^2=\tilde{\epsilon}_{h}+\tilde{\epsilon}_{z}.
\end{align}
Similar to the TKE flux decomposition, the viscous terms above only contain the strain-rate contribution from the velocity gradient tensor since only these can be properly interpreted as dissipation rates. Additional contributions arise involving the rotation rate, and their role is again to redistribute energy without changing the total TKE
\begin{align}
    \tilde{\epsilon}_{r,h} = 2 \nu \tilde{{S}}_{hj}  \tilde{{R}}_{hj},\;
    \tilde{\epsilon}_{r,z} = 2 \nu \tilde{{S}}_{zj}  \tilde{{R}}_{zj},\;  \, \tilde{\epsilon}_{r,h} + \tilde{\epsilon}_{r,z} = 0.
\end{align}
Further, $\tilde{\chi}$ is the FS APE dissipation rate and $\mathcal{B}^f$ is the FS buoyancy flux
\begin{align}
    \tilde{\chi}=\kappa \|\tilde{\bB}\|^2,\; \mathcal{B}^f\equiv N \tilde{u}_z \tilde{\phi}.
\end{align}
Similarly, the balance of the ensemble-averaged SFS energy for homogeneous, stably stratified turbulence is given by:
\begin{align}
    \partial_t \langle e_{h} \rangle &= \langle P^{sf}_h \rangle + \langle \Pi_{h}\rangle + \langle \Pi_{r,h}\rangle -\langle \varepsilon_{h}\rangle -\langle \varepsilon_{r,h}\rangle +\langle\mathcal{F}^{sf}_{K}\rangle, \label{eq:SFStke_budget_ss1}\\
    \partial_t \langle e_{z} \rangle &= \langle P^{sf}_z \rangle +\langle \Pi_{z}\rangle  + \langle \Pi_{r,z}\rangle -\langle\varepsilon_{z}\rangle -\langle \varepsilon_{r,z}\rangle + \langle \mathcal{B}^{sf}\rangle , \label{eq:SFStke_budget_ss2}\\
    \partial_t \langle e_\phi \rangle &= -\langle \mathcal{B}^{sf}\rangle +\langle \Pi_\phi\rangle -\langle\varepsilon_\phi\rangle, \label{eq:SFStke_budget_ss3}
\end{align}
where, $P^{sf}_h$ and $P^{sf}_z$ are the SFS pressure-redistribution terms,
$\varepsilon_{h}$ and $\varepsilon_{z}$ are the SFS horizontal and vertical TKE dissipation rates,
$\varepsilon_{r,h}$ and $\varepsilon_{r,z}$ are the SFS redistribution by viscous stresses,
$\varepsilon_\phi$ is the SFS APE dissipation rate,
and $\mathcal{B}^{sf}$ is the SFS buoyancy flux. 
These terms are defined as
\begin{align}
    P^{sf}_h &\equiv \frac{1}{\rho_r}\widetilde{ p \nabla_h \bm{\cdot} \bu_h}  - P^{f}_h,\; P^{sf}_z \equiv \frac{1}{\rho_r} \widetilde{p \nabla_z u_z} -  P^{f}_z,\; P^{sf}_h + P^{sf}_z = 0, \label{eq:sfs_PR}\\
    \varepsilon_{h} &\equiv 2 \nu \widetilde{{S}_{hj}{S}_{hj}}  -\tilde{\epsilon}_{h},\;
    \varepsilon_{z} \equiv 2 \nu \widetilde{{S}_{zj} {S}_{zj}} - \tilde{\epsilon}_{z},\label{eq:sfs_TKEdr} \\
    \varepsilon_{r,h} &\equiv 2 \nu \widetilde{{S}_{hj} {R}_{hj}} - \tilde{\epsilon}_{r,h},\;
    \varepsilon_{r,z} \equiv 2 \nu \widetilde{{S}_{hj} {R}_{zj}} - \tilde{\epsilon}_{r,z},\;
    \varepsilon_{r,h}+\varepsilon_{r,z}=0,
    \label{eq:sfs_TKErd}\\
    \varepsilon_\phi &\equiv (\nu/Pr) \left(\widetilde{\|\bB\|^2} - \|\tilde{\bB}\|^2\right),\; 
    \mathcal{B}^{sf}\equiv -N \left( \widetilde{u_z \phi} - \tilde{u}_z \tilde{\phi}\right). \label{eq:sfs_phiterms}
\end{align}

\subsection{Energetic mechanisms conditioned on density gradient alignments}\label{theory:alignment}

It can be shown analytically that the quantities appearing in the equations for the SFS TKE and APE depend upon the fluctuating velocity gradient $\boldsymbol{A}\equiv\boldsymbol{\nabla u}$, the fluctuating rescaled-density gradient $\boldsymbol{B}\equiv\boldsymbol{\nabla}\phi$, their geometric alignments with each other, and the direction of the mean density gradient, i.e. $-\ez$. In particular, using the analytical procedure of \citet{Johnson_2020,Johnson_2021} that assumes $\mathcal{G}_\ell$ is an isotropic Gaussian filter, \citet{Bhattacharjee_2026a} derived 
\begin{align}
\mathcal{B}^{sf} &=  \mathcal{B}^{sf,{\rm l}}+\mathcal{B}^{sf,{\rm nl}} \label{eq:partition_PB},\\
\mathcal{B}^{sf,{\rm l}} &\equiv - \ell^2 N \tilde{\bB} \bm{\cdot} \tilde{\bA}^\top \bm{\cdot} \ez \label{eq:partition_Bl},\\
\mathcal{B}^{sf,{\rm nl}} &\equiv  -N \int_0^{\ell^2}  \boldsymbol{T}^\theta \left( \tilde{\boldsymbol{B}}^{\sqrt{\alpha}}, \widetilde{\boldsymbol{A}^\top}^{\sqrt{\alpha}} \right) \bm{\cdot} \ez\,d \alpha \label{eq:partition_Bnl}.
\end{align}
Here $\theta \equiv \sqrt{\ell^2 - \alpha}$ and $\boldsymbol{T}^\theta(\bm{a},\bm{b})\equiv\widetilde{\bm{a\cdot b}}^\theta - \tilde{\bm{a}}^\theta \bm{\cdot} \tilde{\bm{b}}^\theta$, for arbitrary tensors $\bm{a},\bm{b}$ (and we have shown explicitly the superscript denoting filtering at scales $\theta$, $\sqrt{\alpha}$ in keeping with the definition from \eqref{eq:def_filtering}). 
In keeping with the terminology used in \citet{Johnson_2020,Johnson_2021}, the term $\mathcal{B}^{sf,{\rm l}}$ represents the ``scale-local'' contribution to $\mathcal{B}^{sf}$ which only involves contributions from scales $\geq\ell$, while $ \mathcal{B}^{sf,{\rm nl}} $ is the ``non-local'' contribution which only involves contributions from scales $< \ell$ (the sub-filter scales).

Equations \eqref{eq:partition_PB} - \eqref{eq:partition_Bnl} show that the behaviour of the buoyancy flux $\mathcal{B}^{sf}$ depends on the alignment properties of the filtered tensor fields, $\tilde{\bA}, \tilde{\bB}$, along with $\ez$. Similarly, \citet{Zhang2022} also used the analytical procedure of \citet{Johnson_2020,Johnson_2021} to derive
\begin{align}
\Pi_\phi &=  \Pi_\phi^{{\rm l}}+\Pi_\phi^{{\rm nl}} \label{eq:partition_Pphi},\\
\Pi_\phi^{{\rm l}} &\equiv - \ell^2  \tilde{\bB} \bm{\cdot} \tilde{\bA}^\top \bm{\cdot}\tilde{\bB}   \label{eq:partition_Pphil},\\
\Pi_\phi^{{\rm nl}} &\equiv  -\tilde{\bB}\bm{\cdot}  \int_0^{\ell^2}  \boldsymbol{T}^\theta \left( \tilde{\boldsymbol{B}}^{\sqrt{\alpha}}, \widetilde{\boldsymbol{R}^\top}^{\sqrt{\alpha}} \right)\,d \alpha \label{eq:partition_Pphinl},
\end{align}
where $\bm{R}\equiv (\bA-\bA^\top)/2$ is the rotation-rate tensor. Again, it is clear that the APE flux $\Pi_\phi$ depends on the alignment properties of the filtered tensor fields, $\tilde{\bA}$ and $\tilde{\bB}$. This also suggests an implicit dependence on the alignment of $\tilde{\bB}$ and $\ez$ since, for example, the alignment of $\tilde{\bA}$ and $\tilde{\bB}$ will be very different for the case where the orientation of $\tilde{\bB}$ is isotropically distributed compared with where $\tilde{\bB}$ is strongly aligned with $\pm\ez$. Analogous results can also be derived using the same method for the SFS pressure-strain redistribution $P^{sf}$, the total TKE flux $\Pi_K$, and the total SFS dissipation rates $\varepsilon_{K},\varepsilon_\phi$ (all of which can be decomposed into contributions involving horizontal and vertical fields to obtain $\Pi_{K,h}$ and $\Pi_{K,v}$ etc). What this reveals is that the mechanisms controlling the multiscale distribution of TKE and APE are themselves affected by the alignment properties of the velocity and density gradient fields with the vertical direction. Analyzing how the TKE and APE mechanisms depend on the local gradient alignments can therefore provide insights into the energetics of stratified turbulent flows, in the same way that studying the alignment properties of the vorticity and strain-rate eigenvectors provided important insights into the velocity gradient dynamics and energy cascades in isotropic turbulence \citep{tsinober_book, meneveau_lagrangian_2011,carbone_bragg_2020,Johnson_2020,Johnson_2021}. 

We note that such a connection between the multiscale energetics and the gradient field alignments can only be developed analytically if one analyzes the multiscale energetics using the SFS form of the energy equations, rather than the FS equations which are more commonly used when employing a filtering analysis to investigate the multiscale properties of turbulent flows. This is one key reason why we use the SFS equations rather than the FS equations to analyze the multiscale energetics of stratified turbulence. They do, of course, provide complementary information.

When considering all terms in the TKE and APE equations, it is the alignment properties of $\tilde{\bA}$ and $\tilde{\bB}$, together with the direction $\ez$, that influence the TKE and APE evolution. However, in this study we will focus on the impact of the alignment of $\tilde{\bB}$ with the direction $\ez$. Such an analysis was already conducted for the SFS buoyancy flux in \citet{Bhattacharjee_2026a}; here we extend the analysis to consider all terms in the equations. For this analysis we decompose each ensemble average in \eqref{eq:SFStke_budget_ss1} - \eqref{eq:SFStke_budget_ss3} into an average conditioned on the inner product ${\bm{e}}_{\tilde{B}} \bm{\cdot} \ez \in[-1,+1]$, where ${\bm{e}}_{\tilde{B}}\equiv \tilde{\bB}/\|\tilde{\bB}\|$, and
\begin{align}
\langle a\rangle &=\int_{-1}^{+1} \langle a\rangle_\xi \mathcal{P}(\xi;\ell)\, d\xi,
\end{align}
where $ a(\bm{x},t)$ is an arbitrary field, $\langle\cdot\rangle_\xi$ denotes an average conditioned on $\xi={\bm{e}}_{\tilde{B}} \bm{\cdot} \ez$, and $\mathcal{P}(\xi;\ell)\equiv\langle\delta( \xi- {\bm{e}}_{\tilde{B}} \bm{\cdot} \ez )\rangle$ is the probability density function (PDF) of the sample-space variable $\xi$ which is conjugate to the random variable ${\bm{e}}_{\tilde{B}} \bm{\cdot} \ez$. The variable $\ell$ is explicitly included in the argument of the PDF since ${\bm{e}}_{\tilde{B}}$ depends on the filter length $\ell$ through $\tilde{\bB}$.

By analyzing quantities such as $\langle\Pi_\phi \rangle_\xi$, for example, we can gain insight into how the energetics of stratified turbulence depends on the local alignment of the fluctuating and mean density gradient in the flow. In \S\ref{sec:Res_align_and_stab} we will also consider how these geometric alignment properties of the flow are related to notions of local flow stability and instability at different scales, which are very important for the dynamics and energetics of stratified turbulence.

Finally, we note that terms which vanished from \eqref{eq:SFStke_budget_ss1} - \eqref{eq:SFStke_budget_ss3} (e.g. transport terms) due to having averages that are zero under the assumed symmetries of the flow, would not necessarily have conditional averages that are also zero. Hence a complete analysis of how the mechanisms governing the multiscale TKE and APE depend on the local gradient alignment properties would also have to consider the behavior of these terms. However, to limit the scope of this study, the analysis of these terms will be kept for a future study.

\section{Direct Numerical Simulations}\label{sec:Numerical}

The data sets to be used in the next section are from the simulation campaign of statistically stationary, stably-stratified DNSs first described in \cite{Almalkie_2012}, which have later been examined by \citet{debk15}, \cite{Portwood2016}, \cite{Taylor2019}, \cite{CouchmandBKCC2023}, \cite{Petropoulos2024}, and \cite{Bhattacharjee_2026a}. The DNS for $Pr=1,7$ have been introduced and analyzed in the aforementioned papers, whereas the data sets for $Pr=50$ have not been previously presented. For details of the simulation, see \cite{Almalkie_2012}. 

In the DNS, the unfiltered versions of equations \ref{eq:filteredNSE_boussinesqA} and \ref{eq:filteredNSE_boussinesqB} are solved numerically in a triply periodic domain with horizontal and vertical dimensions $\mathcal{L}_h=2\pi$ and 
$\mathcal{L}_z =\mathcal{L}_h/\text{AR}$ (AR is the aspect ratio), respectively.

The DNS were run with three different Prandtl numbers, $Pr=1,7,50$, and for Froude number $Fr\approx 0.16$ (DNS with $Fr\approx 0.08$ were also conducted, but unsteadiness of the results made them unsuitable for the purposes of our analysis). Note that the definition of $Fr$ we use is smaller by a factor of $2\pi$ relative to the references cited at the start of this section, which we use for consistency with the more traditional definition $Fr\equiv u'/(LN)$ introduced earlier. In each case, a constant activity parameter $Gn \equiv \langle\epsilon \rangle/(\nu N^2) \approx 50$ is maintained. Further details on the parameters from the DNS are given in Table \ref{tab:Parameters}.

\begin{table}
\centering
\captionsetup{width=\linewidth}
\begin{tabular}{c c c c c c c c c} 
$Gn$    &   $Pr$    &   $Fr$     &  AR   &   $n_h$   &   $L/\eta$  & $L/l_{O}$&   $l_{O}/\eta$ &   $\eta/\eta_B$  \\ \\ 
$50$    &   $1$     &   $0.16$   &  $4$  &   $4096$  &   $649.8$   & $27.6$    &   $23.6$       &   $1$             \\ 
$50$    &   $7$     &   $0.16$   &  $4$  &   $8192$  &   $511.4$   & $27.2$    &   $18.8$       &   $\sqrt{7}$      \\ 
$50$    &   $50$    &   $0.16$   &  $4$  &   $28672$ &   $368.9$   & $20.6$    &   $18.0$       &   $\sqrt{50}$     \\ 
\end{tabular}
\caption{Flow parameters in DNS. $Gn\equiv \langle \epsilon \rangle/\nu N^2$ is the activity parameter, $Fr\equiv u'/(LN)$ is the Froude number. AR$=\mathcal{L}_h/\mathcal{L}_z$ is the aspect ratio of the simulation domain, where $\mathcal{L}_h=2 \pi$, $\mathcal{L}_z$ are the lengths of the domain in the horizontal ($h$) and the vertical ($z$) directions, respectively. $n_h$ and $n_z=n_h/\text{AR}$ are the corresponding number of grid points in the horizontal and the vertical directions. 
$L$ is the integral length scale computed as the horizontal autocorrelation length of the horizontal field. $l_{O}$, $\eta$ and $\eta_B$ are the Ozmidov, Kolmogorov and Bachelor length scales, respectively. 
$(\mathcal{L}_h/n_h)/\eta_B<2.5$ for all simulations.
\label{tab:Parameters}}
\end{table}

Computing statistical quantities through a combination of space and time averages is extremely challenging for these data sets due to the enormous number of grid points used in the DNS, especially for the higher $Pr$ cases. Due to this, the statistical results in the following section are based on taking spatial averages over the entire domain but for a single snapshot. Unfortunately, performing averages over many snapshots is not likely to be feasible any time soon given available computing resources. For smaller-scale quantities, this is not necessarily problematic because for the DNS parameters being considered, there are many small-scales in the domain over which the averaging is being performed. It could introduce uncertainties for larger-scale analysis, however. One practical way of testing for convergence of the results involves computing the terms in the TKE and APE energy budgets as a function of scale, and measuring the ``residuals'' which correspond to the spatial averages of the time derivative terms in the TKE and APE equations (all terms involving averages of spatial gradients vanish identically due to periodicity of the domain). The results in figure \ref{fig:budget_TKPE} (which will be discussed later) show that for $Pr=1,7$ the residual terms are small compared to the other non-zero terms in the equations. For $Pr=50$, the residual term is subleading in the equations, but is comparable to the magnitude of the buoyancy flux term in the APE equation at small-scales.

The post-processing computations were performed in mixed single-point precision to alleviate the computational bottleneck, which is the MPI communication used when computing Fourier-transforms. However, the accuracy of mixed-precision statistics have been verified with fully double-precision computations for the simulations with $Pr=1$. For the post-processing analysis, an isotropic Gaussian filter was used for consistency with the theoretical results presented in \S \ref{theory:alignment}. 
We have chosen logarithmically spaced filter-scales 
such that the set of $\ell/\eta$ values used in the analysis is constant across all simulations. 

\section{Results}\label{sec:Results}





\subsection{Unconditioned energy budgets}

\begin{figure}
\centering
\includegraphics[trim = {0mm 85mm 125mm 19mm}, scale=0.75,clip]{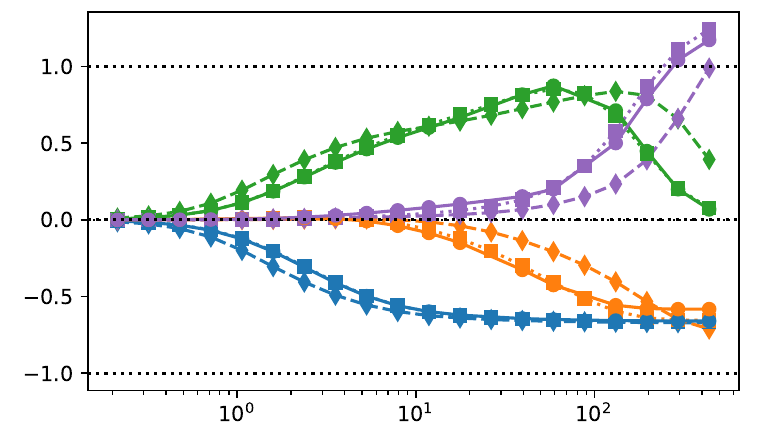}
\put(-290,150){\footnotesize (a)}
\put(-275,40){\rotatebox{90}{{\footnotesize $\langle P_h\rangle, \langle \Pi_{K,h}\rangle,\langle \varepsilon_{K,h} \rangle, \langle R\rangle $}}}\\
\includegraphics[trim = {125.46mm 85mm 0mm 19mm}, scale=0.75,clip]{Figures_Budget/Pr1750Fr1Gn50SGS_balance.pdf}
\put(-290,150){\footnotesize (b)}
\put(-275,17){\rotatebox{90}{{\footnotesize $\langle P_z\rangle, \langle \Pi_{K,z}\rangle,\langle \varepsilon_{K,z} \rangle, \langle \mathcal{B}^{sf}\rangle, \langle R\rangle$}}}\\ 
\includegraphics[trim = {0mm 0mm 125mm 101mm}, scale=0.75,clip]{Figures_Budget/Pr1750Fr1Gn50SGS_balance.pdf}
\put(-290,150){\footnotesize (c)}
\put(-275,40){\rotatebox{90}{{\footnotesize $\langle \Pi_{\phi}\rangle,\langle \varepsilon_{\phi} \rangle, -\langle \mathcal{B}^{sf}\rangle, \langle R\rangle$}}}
\put(-132,0){\footnotesize $\ell/\eta$}\\
\flushleft
\includegraphics[trim = {10mm 20mm 0mm 5mm}, scale=0.6,clip]{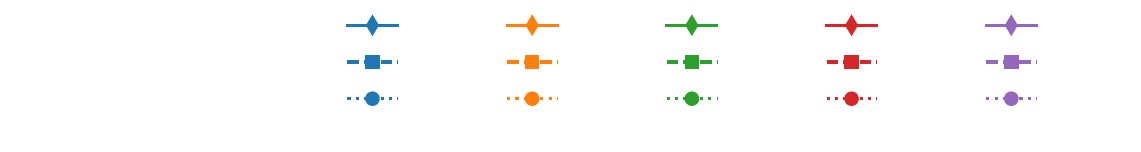}
\put(-250,33.5){\footnotesize $Pr_1$:}
\put(-250,24){\footnotesize $Pr_2$:}
\put(-250,14){\footnotesize $Pr_3$:}
\put(-210,8){\rotatebox{90}{\footnotesize dissipation}}
\put(-165,13){\rotatebox{90}{\footnotesize pressure-}}
\put(-155, 4){\rotatebox{90}{\footnotesize redistribution}}
\put(-120,11){\rotatebox{90}{\footnotesize scalewise}}
\put(-110,21){\rotatebox{90}{\footnotesize flux}}
\put( -70,10){\rotatebox{90}{\footnotesize buoyancy}}
\put( -60,21){\rotatebox{90}{\footnotesize flux}}
\put( -25,12){\rotatebox{90}{\footnotesize forcing +}}
\put( -15,12){\rotatebox{90}{\footnotesize unsteady}}

\hfill
\captionsetup{width=\linewidth} 
\caption{(a) Horizontal and (b) vertical sub-grid-scale (SFS) turbulent kinetic energy balance (normalized by $\langle \epsilon \rangle$) as a function of scale $\ell/\eta$, for $Pr_1=1$ (dashed), $Pr_2=7$ (dash-dotted with squared) and $Pr_3=50$ (solid with circle). (c) SFS Turbulent potential energy (APE) balance (normalized by $\langle \chi \rangle$) across scale.
\label{fig:budget_TKPE}}
\end{figure}

We first consider the behaviour of the ensemble averaged terms appearing in the SFS energy  budget equations \eqref{eq:SFStke_budget_ss1} - \eqref{eq:SFStke_budget_ss3}. In Figure \ref{fig:budget_TKPE} we show (a) the horizontal and (b) the vertical components of the SFS TKE balance as a function of the filter length $\ell/\eta$, for $Pr_1=1$ (dashed with diamond), $Pr_2=7$ (dash-dotted with square) and $Pr_3=50$ (solid with circle). Figure \ref{fig:budget_TKPE} (c) shows the SFS APE balance as a function of the filter length $\ell/\eta$. The purple lines in panel (a) correspond to the forcing and the residual (unsteadiness), whereas those in (b) and (c) correspond to the unsteadiness terms (since only the horizontal TKE is forced). The forcing and unsteadiness terms are not computed directly; rather, they are the residual of the sum of the terms in the right-hand sides of \eqref{eq:SFStke_budget_ss1} - \eqref{eq:SFStke_budget_ss3}. The spatial derivative terms which were assumed to be zero in the theoretical analysis due to homogeneity are in fact identically zero for the present flow due to the use of periodic boundary conditions and spatial averages to evaluate ensemble averages.

In the limit $\ell/\eta \to\infty$ the balances reduce to the energy balances for the unfiltered fields. The results in figure \ref{fig:budget_TKPE} (a) reflect this, showing that for the horizontal TKE there is an approach to a balance between the energy injection term (forcing), the TKE dissipation rate, and the pressure redistribution for the largest $\ell/\eta$. The dissipation and redistribution terms are almost equal at the largest scales, indicating that the flow is strongly stratified. While the dissipation term remains constant until $\ell/\eta\leq O(10)$ (reflecting the fact that dissipation in the flow is predominantly a small-scale quantity), the redistribution term steadily reduces as $\ell/\eta$ is reduced, and its role is negligible at $\ell/\eta\leq O(10)=O(l_{O}/\eta)$, as expected. However, its decay with reducing $\ell/\eta$ is considerably slower than that for the forcing term. This indicates that while the forcing is confined mainly to the largest scales of the flow, the redistribution term involves contributions from scales across the range $O(l_{O})\leq \ell\leq O(L)$. The pressure-redistribution mechanism is therefore not exclusively a large-scale effect. A significant implication of this is that a stratified inertial range with $\langle\Pi_{K,h}\rangle\approx\langle \epsilon_{h} \rangle =$ constant cannot emerge. In particular, since the redistribution term $\langle P_h \rangle$ is negative and reduces in magnitude with decreasing $\ell$, and since $\langle \varepsilon_{K,h} \rangle$ is approximately constant for $\ell\geq O(l_{O})$, then the balance $\langle\Pi_{K,h}\rangle\approx\langle \varepsilon_{K,h} \rangle-\langle P_h \rangle$ at scales where the forcing is negligible implies that $\langle\Pi_{K,h}\rangle$ must be a decreasing function of decreasing $\ell$. This is precisely what the results in figure \ref{fig:budget_TKPE} show. We note, however, that previous analysis of stratified turbulence using Fourier analysis has provided evidence of a constant horizontal TKE flux and associated horizontal TKE spectrum with scaling $\propto k_h^{-5/3}$ (where $k_h$ is the horizontal wavenumber) in the stratified inertial range, e.g. \citet{Lindborg2006,Almalkie_2012}. Future investigation is needed to understand whether the reason that we do not observe such a constant TKE flux regime is due to differences in $Gn, Fr, Pr$ in our DNS compared to those studies, or if it is due to our use of an isotropic, low pass filtering operator to analyze the multiscale properties of the flow rather than Fourier analysis.

For the vertical TKE balance, the pressure-redistribution term is the source term at large scales, and this term decays faster for $Pr=1$ than for the cases with $Pr>1$. At the large scales, the redistribution term is balanced by the dissipation and buoyancy terms. The dissipation term remains constant until $\ell/\eta\leq O(10)$, again reflecting the fact that dissipation in the flow is a small-scale quantity. For $\ell >O(l_{O})$, the buoyancy flux $\langle \mathcal{B}^{sf} \rangle$ is negative, corresponding to the transfer of vertical TKE to APE. The $Pr$ dependence of $\langle \mathcal{B}^{sf} \rangle$ largely reflects that of the redistribution term.

At scales smaller than the Ozmidov scale $\ell < l_{O}$ (for the current data-sets $l_{O}/\eta \approx 20$; see table \ref{tab:Parameters} for exact values) the APE energy balance depends strongly on $Pr$.  For $Pr=1$, the buoyancy flux becomes sub-leading at $\ell < l_{O}$, which leads to an approximate cascade balance $\langle \Pi_\phi \rangle \approx \langle \varepsilon_\phi \rangle \approx \langle \chi \rangle$. However, the range over which there is an approximate cascade is small at the $Gn$ value considered. For $Pr=7,50$, however, $\langle \mathcal{B}^{sf} \rangle$ switches sign below $\ell < l_{O}$, i.e. a reverse buoyancy flux is observed with $\langle \mathcal{B}^{sf} \rangle>0$, as discussed in \citet{Bhattacharjee_2026a}. Note that this sign-reversal also occurs for $Pr=1$, but the magnitude is very small, unlike for higher $Pr$.
(We remind the reader again that non-stationarity of the flow is quite significant for the $Pr=50$ simulations and the APE balance. Hence, the exact values of the flux for this case should be viewed with some caution). The flux reversal implies that $\langle \mathcal{B}^{sf} \rangle$ acts as a sink for APE and as a source for vertical TKE at smaller scales, opposite to its role for $\ell>l_{O}$. For $\ell\leq O(\eta)$, the magnitude of $\langle\mathcal{B}^{sf}\rangle/\langle\chi\rangle$ increases with increasing $Pr$, consistent with the analysis in \citet{Bhattacharjee_2026a}. At scales larger than this but $\ell<l_{O}$, the magnitude of $\langle\mathcal{B}^{sf}\rangle/\langle\chi\rangle$ reduces in going from $Pr=7$ to $Pr=50$. 

An important implication of the reverse buoyancy flux and the fact that $\langle\mathcal{B}^{sf}\rangle/\langle\chi\rangle$ takes on non-negligible values for $Pr>1$ is that a constant down-scale APE cascade at scales $\ell < l_{O}$ cannot occur. This is because while APE is being on average transferred downscale via $\langle\Pi_\phi\rangle$, some of the energy leaks out to the vertical TKE field due to the reverse buoyancy flux, rendering a constant flux impossible. This is very different from what is expected based on the standard theory of stably stratified turbulence according to which the SFS field $\phi-\tilde{\phi}$ should behave as a passive scalar when $\ell \ll l_{O}$, with a constant APE flux $\langle\Pi_\phi\rangle\approx \langle\chi\rangle$. The breakdown of the APE cascade when $Pr\gg 1$ is not clear in our results because $Gn$ is not large enough to see a clear, extended cascade region for $Pr=1$, and therefore departures from this for $Pr\gg1$ cannot be identified.

Another interesting observation based on figure \ref{fig:budget_TKPE} is that almost none of the energy that the buoyancy flux feeds back into the small-scale vertical TKE is shared by the pressure-velocity fluctuations with the horizontal TKE.
This is consistent with the observation by \cite{Okino2019} that only the vertical TKE shows oscillatory behaviour in decaying stratified turbulence due to the oscillating buoyancy flux; the horizontal kinetic energy decays monotonically.

\subsection{Alignment PDFs}

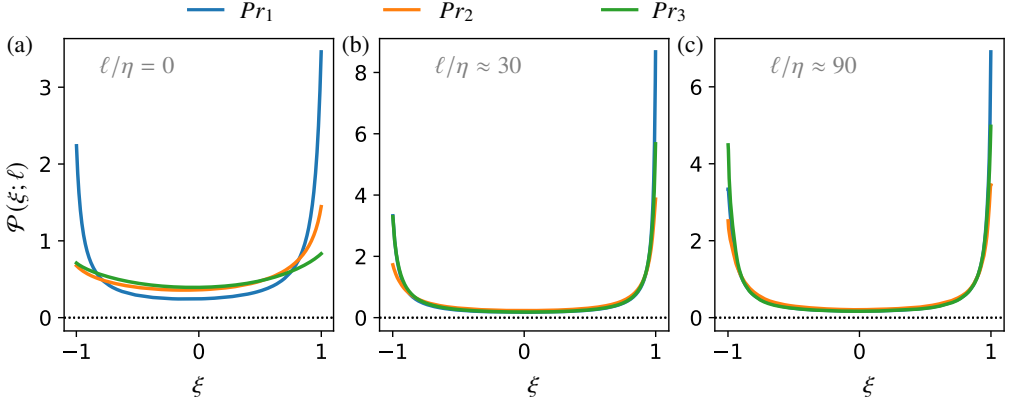
\begin{figure}
\centering
\includegraphics[trim = {15cm 7.5cm 2cm 1cm}, scale=0.4,clip]{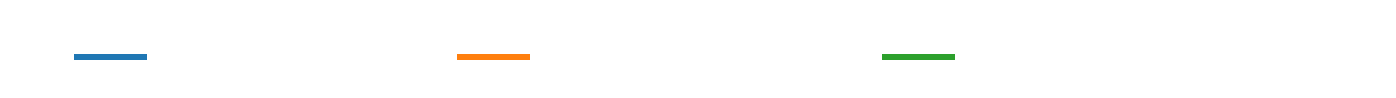}
\put(-235,5){{\footnotesize $Pr_1$}}
\put(-160,5){{\footnotesize $Pr_2$}}
\put(- 80,5){{\footnotesize $Pr_3$}}\\
\includegraphics[trim = {3mm 0mm 5mm 16mm}, scale=0.65,clip]{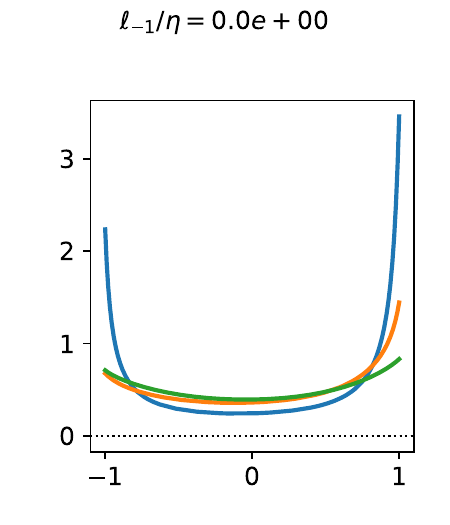}
\put(-125,128){{\footnotesize (a)}}
\put(-125,60){\rotatebox{90}{{\footnotesize $\mathcal{P}(\xi;\ell)$}}}
\put(-90,120){\textcolor{gray}{\footnotesize $\ell/\eta=0$}}
\put(-55,0){{\footnotesize $\xi$}}
\includegraphics[trim = {3mm 0mm 5mm 16mm}, scale=0.65,clip]{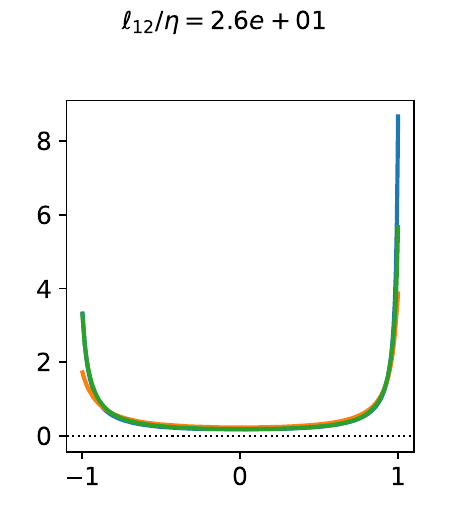}
\put(-125,128){{\footnotesize (b)}}
\put(-90,120){\textcolor{gray}{\footnotesize $\ell/\eta\approx 30$}}
\put(-55,0){{\footnotesize $\xi$}}
\includegraphics[trim = {3mm 0mm 5mm 16mm}, scale=0.65,clip]{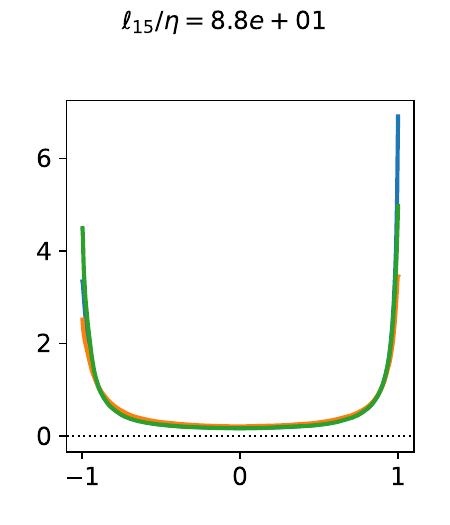}
\put(-125,128){{\footnotesize (c)}}
\put(-90,120){\textcolor{gray}{\footnotesize $\ell/\eta\approx 90$}}
\put(-55,0){{\footnotesize $\xi$}}
\captionsetup{width=\linewidth} 
\caption{
Plots of $\mathcal{P}(\xi;\ell)$, the PDF of the inner product of ${\bm{e}}_{\tilde{\bB}}$ with $\ez$, for $Pr_1=1$ (blue), $Pr_2=7$ (green), and $Pr_3=50$ (orange) at $Fr \approx 0.16$ for (a) $\ell/\eta = 0$, (b) $\ell/\eta \approx 30$, and (c) $\ell/\eta \approx 90$. \label{fig:alignment_Bg} }
\end{figure}
Before analyzing the energetic mechanisms conditioned on the density gradient alignments, we first consider the PDFs of the alignments themselves in order to gain an understanding of their behaviour. Figure \ref{fig:alignment_Bg} shows plots of $\mathcal{P}(\xi;\ell)\equiv\langle\delta( \xi- {\bm{e}}_{\tilde{B}} \bm{\cdot} \ez )\rangle$, the PDF of the inner product of ${\bm{e}}_{\tilde{\bB}}$ with $\ez$ at scales (a) $\ell/\eta=0, (b) \ell/\eta \approx 30$, and (c) $\ell/\eta \approx 90$. The results for the different $Pr$ are shown in blue ($Pr_1=1$), orange ($Pr_2=7$), and green ($Pr_3=50$). For the unfiltered case $\ell/\eta=0$, a strong preference for $\xi>0$ is observed. This corresponds to the formation of ramp-cliff structures in the flow, which are associated with both ${\tilde{B}_z}\equiv\bm{e}_z\bm{\cdot}{\tilde{\bB}}$ being negatively skewed, and the probability of ${\tilde{B_z}}>0$ exceeding that of ${\tilde{B_z}}<0$ \citep{BraggdBKops_2024}. The preferential alignments significantly weaken with increasing $Pr$ for $\ell/\eta =0$, consistent with ramp-cliffs becoming weaker with increasing $Pr$ in the passive scalar regime \citep{Buaria_2021}. Comparing the results to those for a passive scalar in \citet{BraggdBKops_2024} we see that the asymmetry of $\mathcal{P}(\xi;\ell)$ is much stronger for our stratified flows than it is in a flow where density is a passive scalar. 

\begin{figure}
\centering
\includegraphics[trim = {10cm 8cm 2cm 1cm}, scale=0.4,clip]{Legends/LegendA.pdf}
\put(-235,0){{\footnotesize $Pr_1$}}
\put(-160,0){{\footnotesize $Pr_2$}}
\put(- 80,0){{\footnotesize $Pr_3$}}
\flushright
\includegraphics[trim = {0mm 20mm 0mm 16mm}, scale=0.605,clip]{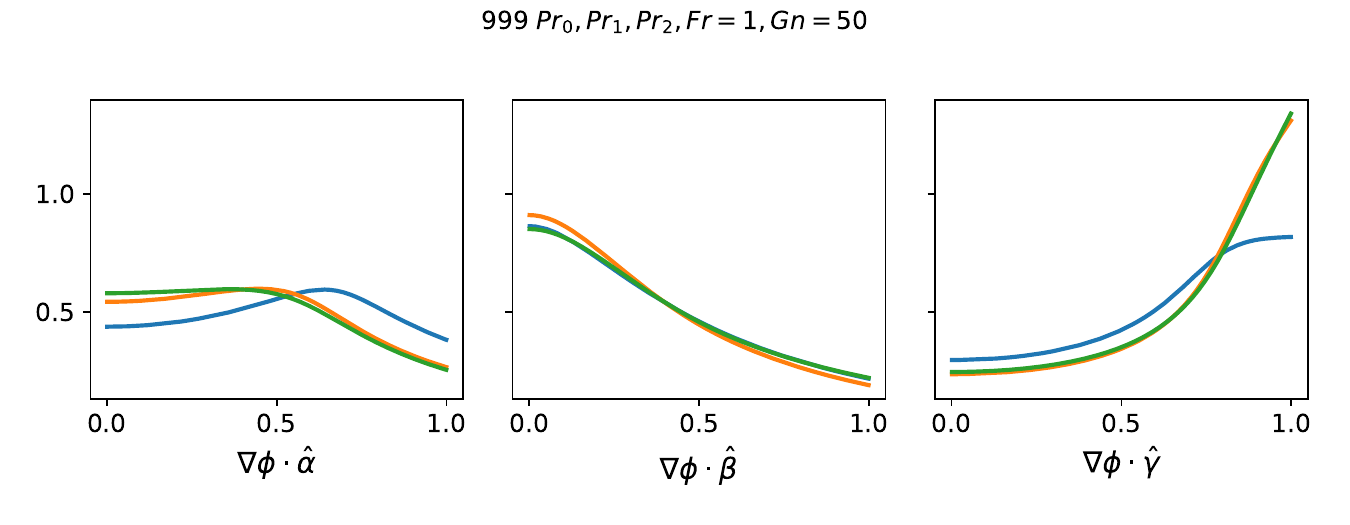}
\put(-395,28){\rotatebox{90}{{\footnotesize $\varphi^B(\xi_i;\ell)$}}}
\put(-390,80){\footnotesize (a)}
\put(-82, 62){\textcolor{gray}{\footnotesize $\ell/\eta = 0$}}\\
\includegraphics[trim = {0mm 15mm 0mm 10mm}, scale=0.605,clip]{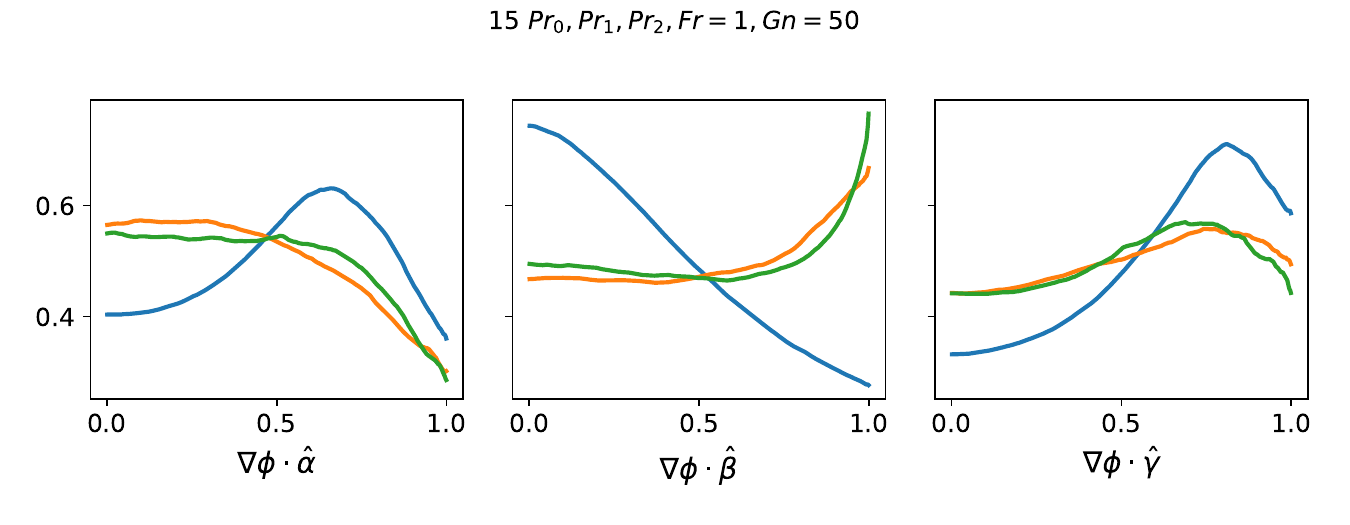}
\put(-390,80){{\footnotesize (b)}}
\put(-395,28){\rotatebox{90}{{\footnotesize $\varphi^B(\xi_i;\ell)$}}}
\put(-310,-10){\footnotesize $\xi_\alpha$}
\put(-190,-10){\footnotesize $\xi_\beta$}
\put(-72, -10){\footnotesize $\xi_\gamma$}
\put(-82, 82){\textcolor{gray}{\footnotesize $\ell/\eta \approx 90$}}
\captionsetup{width=\linewidth} 
\caption{Plots of $\varphi^B(\xi_i;\ell)$, the PDF of inner product of ${\bm{e}}_{\tilde{\bB}}$ with the strain-rate eigenvectors ${\tilde{\bm{\alpha}}}$, ${\tilde{\bm{\beta}}}$, and ${\tilde{\bm{\gamma}}}$ for (a) the unfiltered field i.e. $\ell/\eta = 0$ and (b) for $\ell/\eta \approx 90$. 
The curves correspond to $Pr_1=1$ (blue), $Pr_2=7$ (orange), and $Pr_3=50$ (green). 
\label{fig:Ealign_B} }
\end{figure}

For a given $Pr$, the preferential alignment is stronger for $\ell/\eta\approx 90$ than for $\ell/\eta=0$. This reflects the fact that at larger scales the buoyancy flux plays the key role in supplying APE and for this mechanism to be effective requires significant alignment of ${\bm{e}}_{\tilde{\bB}}$ with $\ez$ \citep{Bhattacharjee_2026a}. At smaller scales, where the APE flux $\Pi_\phi$ plays the dominant role rather than the buoyancy flux, strong alignment of ${\bm{e}}_{\tilde{\bB}}$ is required with respect to the compressive strain-rate eigendirection rather than $\ez$ (see Fig.~\ref{fig:Ealign_B}), although significant alignment with $\ez$ still persists at these scales. At larger scales where the alignment of ${\bm{e}}_{\tilde{\bB}}$ with $\ez$ is strong, the sensitivity to $Pr$ is weak except for events where $|\xi|\approx 1$. The preferential alignments for $\ell/\eta\approx 30$ are however stronger than for $\ell/\eta\approx 90$, suggesting a non-monotonic dependence on scale. This may be due to a mutually enhancing effect of inertia and buoyancy forces on the alignments at scales $\ell=O(\ell_O)$.

An important point that follows from these results is that ramp-cliff structures do not weaken with increasing $Pr$, at least when considered as multiscale objects. Although the skewness of the unfiltered gradient ${{B_z}}$ vanishes in the limit $Pr\to \infty$ \citep{Buaria2021a,Shete2022}, that of ${\tilde{B_z}}$ will not at larger scales. Indeed, ramp-cliff structures will always be present at some range of scales regardless of $Pr$ because it is the mean scalar gradient after all that produces the amplification of ${\tilde{\bB}}$ \citep{BraggdBKops_2024}, and this will always lead to some preferential alignment of ${\bm{e}}_{\tilde{\bB}}$ with $\ez$ at some scales, regardless of $Pr$. 

Although as discussed earlier, we will focus on how the TKE and APE energetic mechanisms depend on the local alignment between $\tilde{\bB}$ and $\ez$, the terms in the TKE and APE equations are more generally impacted by alignments involving $\tilde{\bA}$ as well as $\tilde{\bB}$ and $\ez$. It is therefore instructive to consider some of the additional alignment properties involving $\tilde{\bA}$ and its decomposition into the strain-rate $\tilde{\bS}$ and vorticity $\tilde{\boldsymbol{\omega}}$, which provide insight into the geometry of the flow and will prove helpful when interpreting some of the later results.

Figure \ref{fig:Ealign_B} shows the probability-density function (PDF) of the inner product of ${\bm{e}}_{\tilde{\bB}}$ and the eigenvectors of the filtered strain-rate tensor ${\tilde{\bm{S}}}$, namely ${\tilde{\bm{\alpha}}}$, ${\tilde{\bm{\beta}}}$, and ${\tilde{\bm{\gamma}}}$, which are normalized and ordered (according to decreasing value of their associated eigenvalues). This PDF is denoted by $\varphi^B(\xi_i;\ell)$, where $i=\{\alpha,\beta,\gamma\}$, and we show results for (a) $\ell/\eta = 0$ and (b) $\ell/\eta \approx 90$. For the unfiltered field $\ell/\eta = 0$, the invariant $\langle{\bB} \bm{\cdot} {\bA}^\top \bm{\cdot}{\bB}\rangle$ is negative and is the dominant production mechanism for $\langle\|\bB\|^2\rangle$ \citep{BraggdBKops_2024}, requiring that ${\bm{e}}_{\tilde{\bB}}$ preferentially aligns with the compressive strain-rate eigen-direction ${\bm{\gamma}}$. This is what the results in Figure \ref{fig:Ealign_B} show, with a relatively high probability of the strongly aligned configuration $|\xi_\gamma|\approx 1$. The strongest misalignment is with respect to the intermediate eigen-direction ${\bm{\beta}}$. The alignment of ${\bm{e}}_{\tilde{\bB}}$ with ${\bm{\gamma}}$ becomes much stronger as $Pr$ is increased from $Pr=1$ to $Pr=7$, but saturates from $Pr=7$ to $Pr=50$. The enhancement of the probability of events with $|\xi_\gamma|\approx 1$ as $Pr$ is increased occurs together with the decrease of the probability of events with $|\xi_\alpha|\approx 1$ as $Pr$ is increased, since $Pr$ has a non-monotonic effect on the PDF of $\xi_\beta$. The enhancement of the probability of events with $|\xi_\gamma|\approx 1$ as $Pr$ is increased may be due to the reduced effect of the diffusion term in the equation for $\bB$ as $Pr$ is increased, since this term will in general be misaligned with the term $- {\bA}^\top \bm{\cdot}{\bB} $ that tends to produce strong alignment of ${\bm{e}}_{\tilde{\bB}}$ with ${\tilde{\bm{\gamma}}}$.

At the larger scale $\ell/\eta \approx 90$, the results show that the strong preferential alignment for $|\xi_\gamma|\approx 1$ gives way to a new alignment behavior. At this larger scale and for $Pr=1$, the PDFs of $\xi_\alpha$ and $\xi_\gamma$ become similar, with the most probable configurations being $|\xi_\alpha|\approx 1/\sqrt{2}$ and $|\xi_\gamma|\approx 1/\sqrt{2}$. For $Pr=1$ the strong misalignment with ${\bm{\beta}}$ persists at this scale. The behaviour for $Pr=7,50$ is however quite different in general from that for $Pr=1$ at $\ell/\eta \approx 90$. The PDF for $\xi_\gamma$ is similar for all three $Pr$ numbers, showing a preferential alignment with $|\xi_\gamma|\approx 1/\sqrt{2}$. However, while the PDF for $\xi_\alpha$ shows a clear preference for $|\xi_\alpha|\approx 1/\sqrt{2}$ when $Pr=1$, there is an approximately uniform PDF over the range $\xi_\alpha\lesssim 0.5$ when $Pr=7,50$. Most striking however is that while the PDF for $\xi_\beta$ shows a strong preference for $\xi_\beta\approx 0$ when $Pr=1$, there is a strong preference for $\xi_\beta\approx 1$ when $Pr=7,50$. Further insight into this striking change with increasing $Pr$ will be given momentarily.

\begin{figure}
\centering
\includegraphics[trim = {10cm 8cm 2cm 1cm}, scale=0.4,clip]{Legends/LegendA.pdf}
\put(-235,0){{\footnotesize $Pr_1$}}
\put(-160,0){{\footnotesize $Pr_2$}}
\put(- 80,0){{\footnotesize $Pr_3$}}\\
\flushright
\includegraphics[trim = {0mm 20mm 0mm 16mm}, scale=0.605,clip]{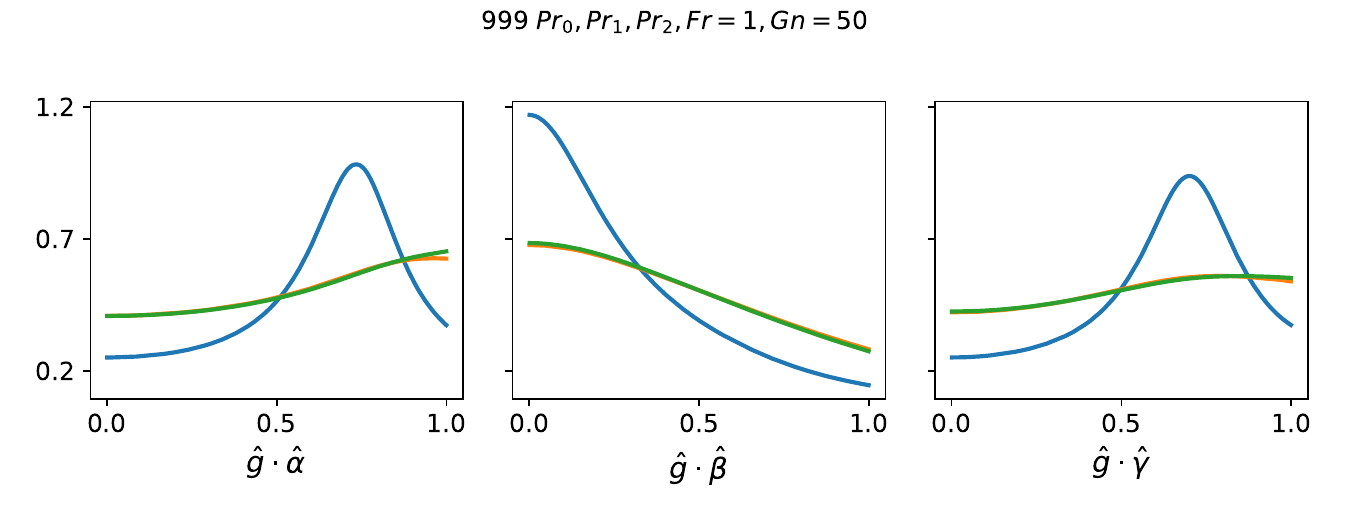}
\put(-395,28){\rotatebox{90}{{\footnotesize $\varphi^z(\xi_i;\ell)$}}}
\put(-395,80){\footnotesize (a)}
\put(-112, 75){\textcolor{gray}{\footnotesize $\ell/\eta = 0$}}\\
\includegraphics[trim = {0mm 15mm 0mm 10mm}, scale=0.605,clip]{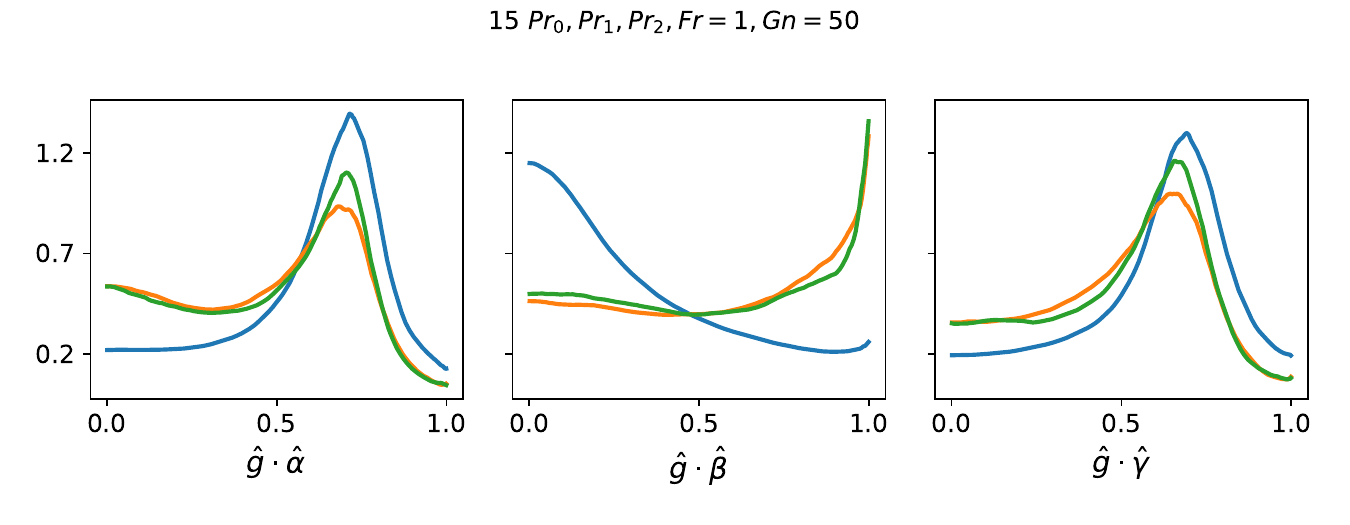}
\put(-395,28){\rotatebox{90}{{\footnotesize $\varphi^z(\xi_i;\ell)$}}}
\put(-395,80){{\footnotesize (b)}}
\put(-310,-10){\footnotesize $\xi_\alpha$}
\put(-190,-10){\footnotesize $\xi_\beta$}
\put(-72, -10){\footnotesize $\xi_\gamma$}
\put(-112, 75){\textcolor{gray}{\footnotesize $\ell/\eta \approx 90$}}
\captionsetup{width=\linewidth} 
\caption{Plots of $\varphi^z(\xi_i;\ell)$, the PDF of inner product of $\ez$ with the strain-rate eigenvectors ${\tilde{\bm{\alpha}}}$, ${\tilde{\bm{\beta}}}$, and ${\tilde{\bm{\gamma}}}$ for (a) the unfiltered field i.e. $\ell/\eta = 0$ and (b) for $\ell/\eta \approx 90$. 
The curves correspond to $Pr_1=1$ (blue), $Pr_2=7$ (orange), and $Pr_3=50$ (green).
\label{fig:Ealign_g} }
\end{figure}

Figure \ref{fig:Ealign_g} shows the PDF of the inner product of $\ez$ with ${\tilde{\bm{\alpha}}}$, ${\tilde{\bm{\beta}}}$, and ${\tilde{\bm{\gamma}}}$, denoted by $\varphi^z(\xi_i;\ell)$, where $i=\{\alpha,\beta,\gamma\}$. Results are shown for (a) the unfiltered field $\ell/\eta = 0$ and (b) $\ell/\eta \approx 90$. If the small-scales of the flow were isotropic then these PDFs would be uniform for $\ell/\eta = 0$ because the vectors ${\bm{\alpha}}$, ${\bm{\beta}}$, and ${\bm{\gamma}}$ would be randomly orientated in space. The results for $\ell/\eta = 0$ and $Pr=1$ show that there is significant anisotropy at the small-scales, despite the fact that other metrics such as $\langle e_{K,h}\rangle/\langle e_{K,z}\rangle$ and $\langle \varepsilon_{K,h}\rangle/\langle \varepsilon_{K,z}\rangle$ (not shown) suggest that the small-scale anisotropy is weak for $Pr=1$. However, for $Pr=7,50$ the PDFs are considerably more uniform, suggesting that the contribution from $\bB$ in the equation for the strain-rate tensor $\bm{S}$ is producing an effect that becomes more isotropic as $Pr$ increases. This is consistent with the earlier observation that the PDF of the inner product of ${\bm{e}}_{\tilde{\bB}}$ and $\ez$ becomes more uniform for $\ell/\eta = 0$, meaning that the orientation of ${\bm{e}}_{\tilde{\bB}}$ is becoming increasingly random as $Pr$ is increased.

For $\ell/\eta \approx 90$ the anisotropy is much stronger and the PDFs of $\xi_\alpha, \xi_\gamma$ are qualitatively similar for all $Pr$. The PDF of $\xi_\beta$ is qualitatively similar for $Pr=1$ at $\ell/\eta = 0$ and $\ell/\eta \approx 90$, while it is very different at these two scales for $Pr=7,50$. The alignment behaviour of $\ez$ with respect to ${\tilde{\bm{\alpha}}}$, ${\tilde{\bm{\beta}}}$, and ${\tilde{\bm{\gamma}}}$ for $Pr=1$ is typical of stratified flows with strong shear between layers where the vertical shear of the horizontal velocity fields are the dominant terms in the velocity-gradient tensor (VGT). \cite{Okino2019} used a simple model of the velocity gradient field with only two non-zero entries to explain the preferential orientations of the eigenvectors. In the Cartesian system with basis vectors $\bm{e}_x,\bm{e}_y,\bm{e}_z$, the vectors ${\tilde{\bm{\alpha}}}$, ${\tilde{\bm{\beta}}}$, and ${\tilde{\bm{\gamma}}}$ in the model are written as
\begin{align} 
\tilde{\bm{\alpha}}&=\frac{\nabla_z\tilde{u}}{2\|\tilde{\bm{S}}\|}\boldsymbol{e}_x+\frac{\nabla_z\tilde{v}}{2\|\tilde{\bm{S}}\|}\boldsymbol{e}_y+\frac{1}{\sqrt{2}}\ez,\label{eq:abg_sheardriven_alpha}\\
{\tilde{\bm{\beta}}}&=\frac{\nabla_z\tilde{v}}{\sqrt{2\|\tilde{\bm{S}}\|}}\boldsymbol{e}_x-\frac{\nabla_z\tilde{u}}{\sqrt{2\|\tilde{\bm{S}}\|}}\boldsymbol{e}_y,\label{eq:abg_sheardriven_beta}\\
{\tilde{\bm{\gamma}}}&=-\frac{\nabla_z\tilde{u}}{2\|\tilde{\bm{S}}\|}\boldsymbol{e}_x-\frac{\nabla_z\tilde{v}}{2\|\tilde{\bm{S}}\|}\boldsymbol{e}_y+\frac{1}{\sqrt{2}}\ez,\label{eq:abg_sheardriven_gamma}
\end{align}
which yields $|\bm{e}_z\bm{\cdot} {\tilde{\bm{\alpha}}}|=|\bm{e}_z\bm{\cdot} {\tilde{\bm{\gamma}}}|=1/\sqrt{2}$ and $|\bm{e}_z\bm{\cdot} {\tilde{\bm{\beta}}}|=0$. These values correspond to the locations of the peaks in the PDFs shown in Figure \ref{fig:Ealign_g} for $Pr=1$. Hence, according to this model, these preferential alignments are a consequence of layering in the flow due to stratification, which is associated with vertical velocity gradients that are much larger than horizontal velocity gradients at larger scales in the flow. Standard stratified turbulence theory would lead to the expectation that these preferential alignments would become negligible for $\ell/\eta =0$ when $Gn \gg 1$, because that is the asymptotic condition generally regarded as being required for buoyancy to be negligible at the small-scales \citep{RileyLindborg_2012}. The analysis of \cite{BraggdBKops_2024} suggest that this is not the case, however, and would instead only occur when $ \Lambda_S = O(Pr^{1/2} Gn^{-1/2} \Gamma^{1/2})$ is sufficiently small. As discussed in \cite{BraggdBKops_2024}, when $Pr=O(1)$, $\Gamma= O(1)$, and hence $ \Lambda_S=O(Gn^{-1/2})$. Therefore, in this case we do not merely require $Gn\gg1$, but the more restrictive condition $Gn^{1/2}\gg1$ to be satisfied if the effects of stratification are to be negligible at the small scales (and this distinction is crucial for DNS, for example, where although $Gn$ may be large enough to satisfy $Gn\gg1$, it may not be large enough to satisfy $Gn^{1/2}\gg1$). When $Pr\gg 1$, even if $Gn^{1/2}\gg1$, this does not guarantee $\Lambda_S\ll1$ unless $\Gamma^{1/2}$ becomes sufficiently small in this regime (see \cite{BraggdBKops_2024} for further discussion).

\begin{figure}
\centering
\includegraphics[trim = {0cm 8cm 0cm 0cm}, scale=0.4,clip]{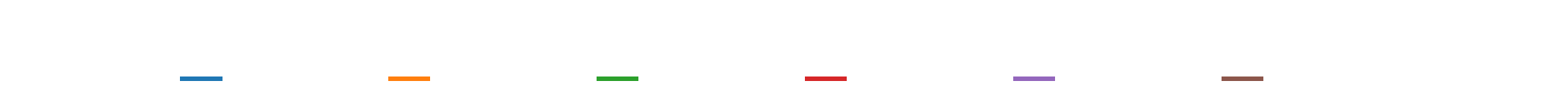}
\put(-341,2){{\footnotesize $\ell_1/\eta$}}
\put(-288,2){{\footnotesize $\ell_2/\eta$}}
\put(-234,2){{\footnotesize $\ell_3/\eta$}}
\put(-181,2){{\footnotesize $\ell_4/\eta$}}
\put(-128,2){{\footnotesize $\ell_5/\eta$}}
\put(-75,2){{\footnotesize $\ell_6/\eta$}}

\flushleft
\includegraphics[trim = {0.2cm 0cm 0cm 1.6cm}, scale=0.6,clip]{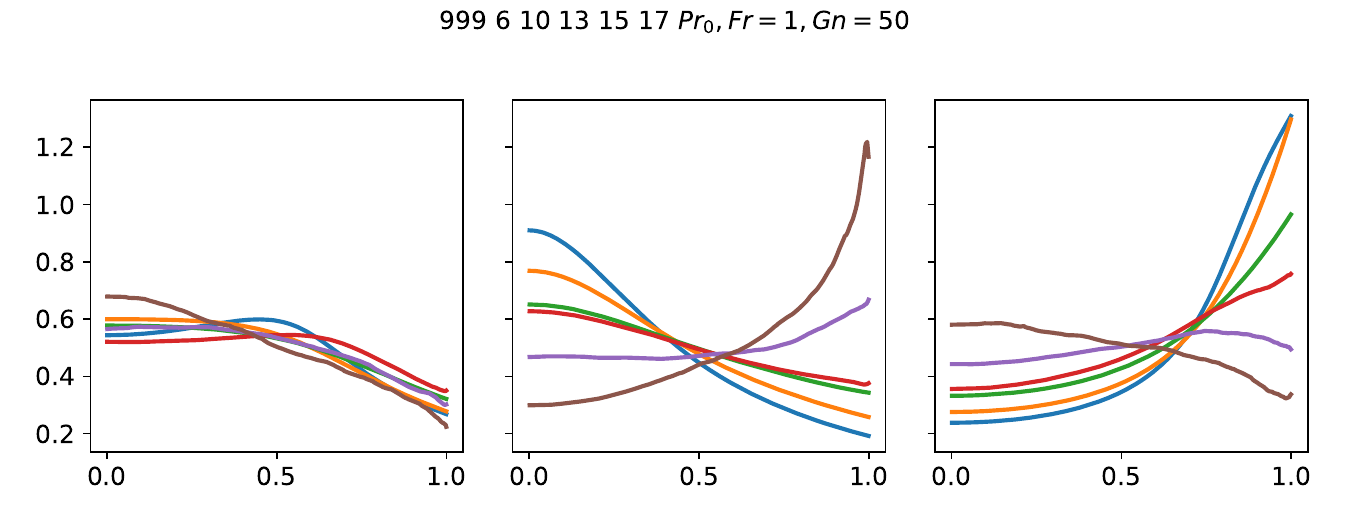}
\put(-395,50){\rotatebox{90}{{\footnotesize $\varphi^B(\xi_i;\ell)$}}}
\put(-385,120){{\footnotesize (a)}}
\put(-255,120){{\footnotesize (b)}}
\put(-132,120){{\footnotesize (c)}}
\put(-315,0){\footnotesize $\xi_\alpha$}
\put(-190,0){\footnotesize $\xi_\beta$}
\put(- 62,0){\footnotesize $\xi_\gamma$}
\captionsetup{width=\linewidth} 
\caption{Plots of $\varphi^B(\xi_i;\ell)$, the PDF of inner product of ${\bm{e}}_{\tilde{\bB}}$ with the strain-rate eigenvectors ${\tilde{\bm{\alpha}}}$, ${\tilde{\bm{\beta}}}$, and ${\tilde{\bm{\gamma}}}$ across different filter-scales: 
 $\ell_1/\eta \approx 0$ (blue), $\ell_2/\eta \approx 2.5$ (orange), $\ell_3/\eta \approx 10$ (green), $\ell_4/\eta \approx 40$ (red), $\ell_5/\eta \approx 90$ (purple), and 
$\ell_6/\eta \approx 200$ (brown) for $Pr_2=7$. 
}
\label{fig:Seigalign_B_cond}
\end{figure}

To understand why the results in Figure \ref{fig:Ealign_B} and Figure \ref{fig:Ealign_g} for the PDFs of $\xi_\beta$ are so different for $Pr=7,50$ from those at $Pr=1$ when $\ell/\eta\approx 90$, in Figure \ref{fig:Seigalign_B_cond} we show the PDF $\varphi^B(\xi_i;\ell)$ for $Pr=7$ but over a wider range of scales $\ell/\eta$. These results show that the shape of $\varphi^B(\xi_i;\ell)$ for both $i=\beta$ and $i=\gamma$ changes dramatically as $\ell/\eta$ is increased, and in particular, it qualitatively changes somewhere in the region $\ell/\eta\in[40,90]$. For $\ell/\eta\leq 40$, the PDF has the same shape shape as it was shown to for $Pr=1$ in Figure \ref{fig:Ealign_B} at scale $\ell/\eta=90$. Therefore, although the results in Figure \ref{fig:Ealign_g} for the PDFs of $\xi_\beta$ look very different for $Pr=7,50$ compared with those for $Pr=1$, it seems that this is simply due to the $Pr$ dependence of the PDF on the lengthscale $\ell/\eta$. In other words, over a particular range of $\ell/\eta$ the results for $Pr=1$ would look qualitatively similar to those for $Pr=7,50$, even though they are not for $\ell/\eta\approx 90$. This is at least in part a reflection of the fact that the PDF is not expected to scale with $\ell/\eta$ since although the eigenvector properties are expected to scale with $\eta$, the properties of the vector ${\bm{e}}_{\tilde{\bB}}$, upon which the PDF also depends, would be expected to scale with $\eta_B$. Exactly how the PDF should scale is unclear given that it depends on both the eigenvectors and ${\bm{e}}_{\tilde{\bB}}$ and their non-trivial alignment properties. We note again that at the larger scales the results may not be statistically converged, and therefore caution should be used in seeking to understand the scale dependence of the PDFs for the larger $Pr$ cases.

\begin{figure}
\centering
\includegraphics[trim = {0cm 8cm 0cm 0cm}, scale=0.4,clip]{Legends/LegendD_horizontal.pdf}
\put(-341,2){{\footnotesize $\ell_1/\eta$}}
\put(-288,2){{\footnotesize $\ell_2/\eta$}}
\put(-234,2){{\footnotesize $\ell_3/\eta$}}
\put(-181,2){{\footnotesize $\ell_4/\eta$}}
\put(-128,2){{\footnotesize $\ell_5/\eta$}}
\put(-75,2){{\footnotesize $\ell_6/\eta$}}

\flushleft
\includegraphics[trim = {0.2cm 0cm 0cm 1.6cm}, scale=0.6,clip]{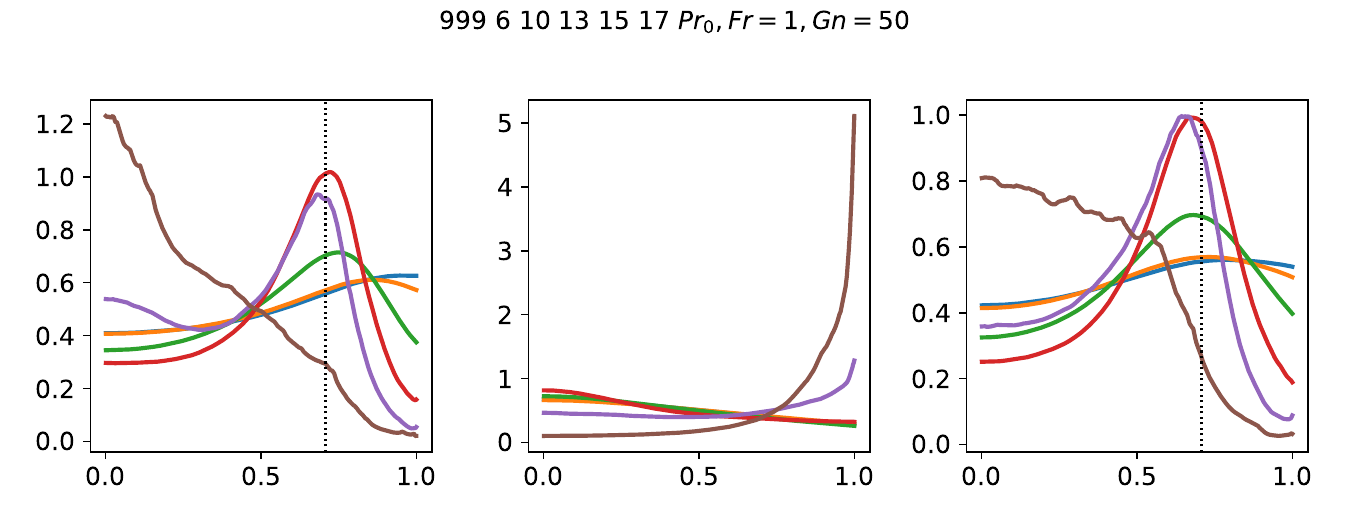}
\put(-395,50){\rotatebox{90}{{\footnotesize $\varphi^z(\xi_i;\ell)$}}}
\put(-385,120){{\footnotesize (a)}}
\put(-255,120){{\footnotesize (b)}}
\put(-132,120){{\footnotesize (c)}}
\put(-315,0){\footnotesize $\xi_\alpha$}
\put(-190,0){\footnotesize $\xi_\beta$}
\put(- 62,0){\footnotesize $\xi_\gamma$}
\captionsetup{width=\linewidth} 
\caption{Plots of $\varphi^z(\xi_i;\ell)$, the PDF of inner product of $\ez$ with the strain-rate eigenvectors ${\tilde{\bm{\alpha}}}$, ${\tilde{\bm{\beta}}}$, and ${\tilde{\bm{\gamma}}}$ across different filter-scales: $\ell_1/\eta \approx 0$ (blue), $\ell_2/\eta \approx 2.5$ (orange), $\ell_3/\eta \approx 10$ (green), $\ell_4/\eta \approx 40$ (red), $\ell_5/\eta \approx 90$ (purple), and 
$\ell_6/\eta \approx 200$ (brown) for $Pr_2=7$. 
}
\label{fig:Seigalign_g_cond}
\end{figure}

It was shown earlier that the alignment properties of ${\bm{e}}_{\tilde{\bB}}$ relative to $\ez$ remain qualitatively the same as $\ell/\eta$ is increased. Therefore, the change in the behaviour of $\varphi^B(\xi_i;\ell)$ for $i=\beta$ as $\ell/\eta$ is increased must be due to a non-trivial dependence of the alignment properties of the eigenvectors relative to $\ez$ as $\ell/\eta$ is increased. This was indicated by the results in Figure \ref{fig:Ealign_g} and is further confirmed by the results in Figure \ref{fig:Seigalign_g_cond} which show the PDF $\varphi^z(\xi_i;\ell)$ for $Pr=7$ over a wider range of scales $\ell/\eta$ than was shown in Figure \ref{fig:Ealign_g}. As noted earlier, the model of \citet{Okino2019} predicts a preferential alignment for states with $|\bm{e}_z\bm{\cdot} {\tilde{\bm{\alpha}}}|=|\bm{e}_z\bm{\cdot} {\tilde{\bm{\gamma}}}|=1/\sqrt{2}$ and $|\bm{e}_z\bm{\cdot} {\tilde{\bm{\beta}}}|=0$. The results in  Figure \ref{fig:Seigalign_g_cond} show that this prediction is confirmed for $i=\alpha,\gamma$ at $\ell/\eta\leq 90$, and for $i=\beta$ at $\ell/\eta\leq 40$. The results for $\ell/\eta\approx 200$ are very different from those predicted by the model, showing a preference for states $|\bm{e}_z\bm{\cdot} {\tilde{\bm{\alpha}}}|=|\bm{e}_z\bm{\cdot} {\tilde{\bm{\gamma}}}|=0$ and $|\bm{e}_z\bm{\cdot} {\tilde{\bm{\beta}}}|=1$. However, we expect that this is not due to the emergent stratified dynamics of the flow, but is rather due to the forcing applied to the flow.

\begin{figure}
\flushright
\includegraphics[trim = {15cm 7.5cm 2cm 1cm}, scale=0.4,clip]{Legends/LegendA.pdf}
\put(-235,5){{\footnotesize $Pr_1$}}
\put(-160,5){{\footnotesize $Pr_2$}}
\put(- 80,5){{\footnotesize $Pr_3$}}\\
\flushright
\includegraphics[trim = {0mm 19mm 0mm 10mm}, scale=0.53,clip]{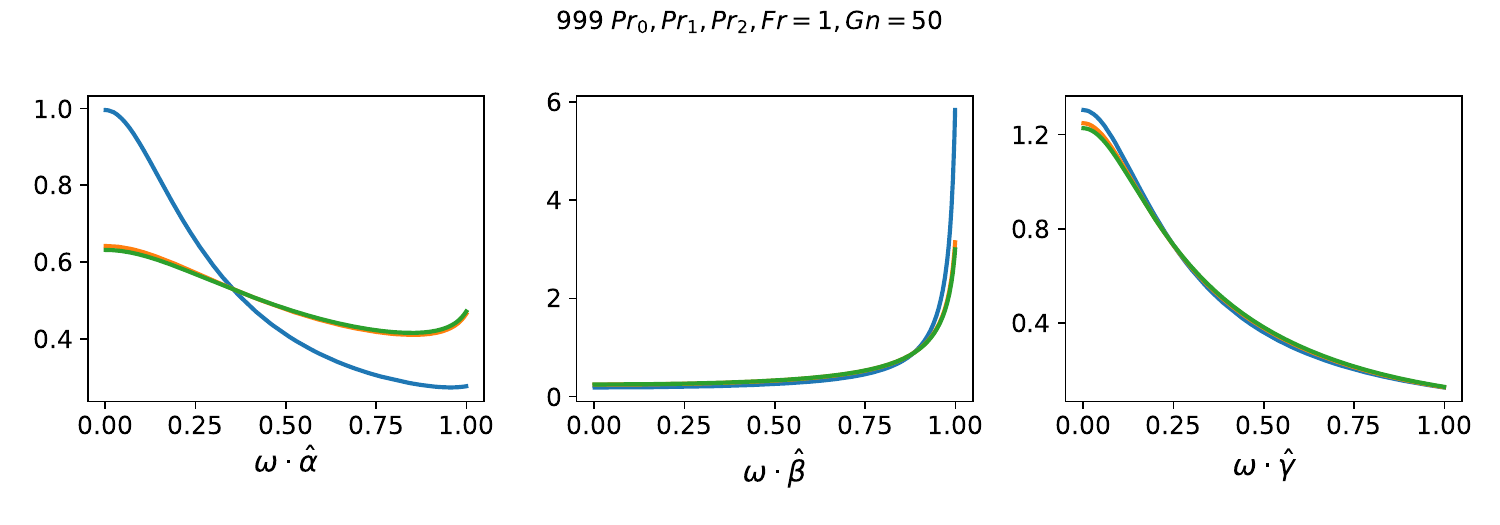}
\put(-385,28){\rotatebox{90}{{\footnotesize $\varrho^\omega(\xi_i;\ell)$}}}
\put(-385,80){\footnotesize (a)}
\put(-70, 65){\textcolor{gray}{\footnotesize $\ell/\eta = 0$}}\\
\includegraphics[trim = {0mm 13mm 0mm 10mm}, scale=0.53,clip]{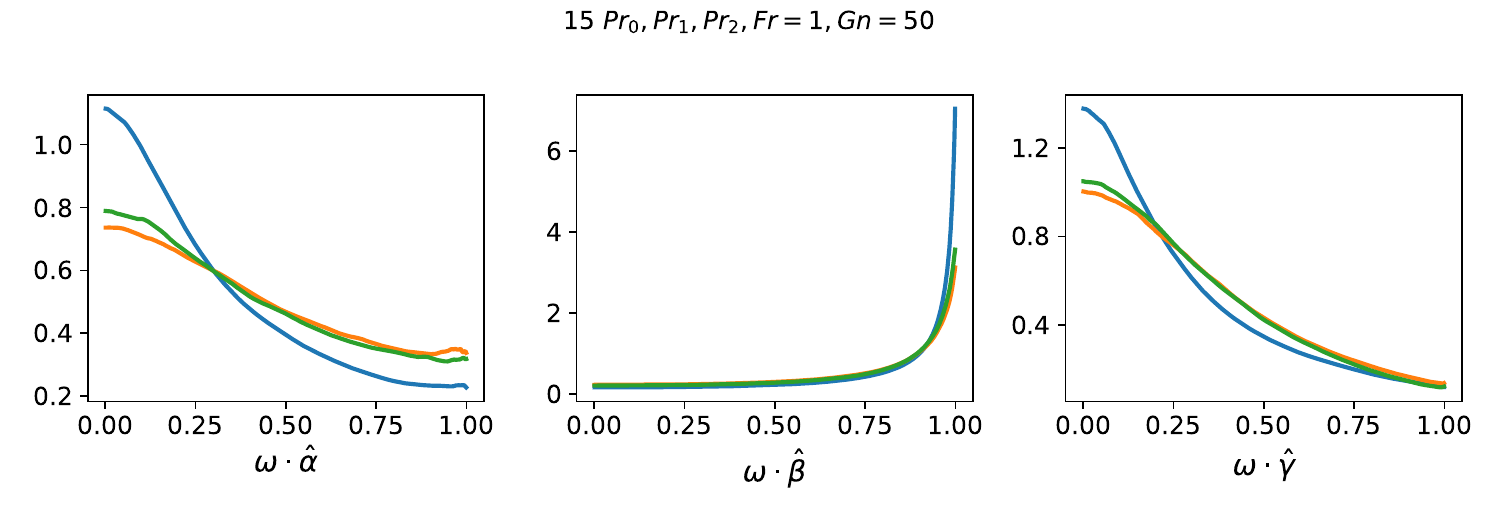}
\put(-385,28){\rotatebox{90}{{\footnotesize $\varrho^\omega(\xi_i;\ell)$}}}
\put(-385,80){{\footnotesize (b)}}
\put(-310,-10){\footnotesize $\xi_\alpha$}
\put(-190,-10){\footnotesize $\xi_\beta$}
\put(-72, -10){\footnotesize $\xi_\gamma$}
\put(-70, 80){\textcolor{gray}{\footnotesize $\ell/\eta \approx 90$}}
\captionsetup{width=\linewidth} 
\caption{Plots of $\varrho^\omega(\xi_i;\ell)$, the PDF of inner product of $\tilde{\bm{\omega}}/\|\tilde{\bm{\omega}}\|$ with ${\tilde{\bm{\alpha}}}$, ${\tilde{\bm{\beta}}}$, and ${\tilde{\bm{\gamma}}}$ for (a) the unfiltered field i.e. $\ell/\eta = 0$ and (b) for $\ell/\eta \approx 90$ (b). The curves correspond to $Pr_1=1$ (blue), $Pr_2=7$ (orange), and $Pr_3=50$ (green).
\label{fig:Ealign_omega} }
\end{figure}

The velocity gradient model of \cite{Okino2019} also predicts that due to strong layering the vorticity should be 
\begin{align} \label{eq:omega_sheardriven}
\tilde{\bm{\omega}}=-\nabla_z\tilde{v}\boldsymbol{e}_x+\nabla_z\tilde{u}\boldsymbol{e}_y.
\end{align}
Equations \ref{eq:abg_sheardriven_beta} and \ref{eq:omega_sheardriven} predict that $|{\tilde{\bm{\beta}}} \bm{\cdot} {\tilde{\bm{\omega}}}| = 1$ i.e. the intermediate eigenvector should be aligned with the vorticity vector. Figure \ref{fig:Ealign_omega} shows results for the PDF of the inner product between $\tilde{\bm{\omega}}/\|\tilde{\bm{\omega}}\|$ and ${\tilde{\bm{\alpha}}}$, ${\tilde{\bm{\beta}}}$, and ${\tilde{\bm{\gamma}}}$, denoted by $\varrho^\omega(\xi_i;\ell)$, where $i=\{\alpha,\beta,\gamma\}$ for $\ell/\eta=0$ (a) and $\ell/\eta\approx 90$ (b). The results show strong alignment of ${\bm{\tilde{\omega}}}$ with ${\tilde{\bm{\beta}}}$, in agreement with equations \ref{eq:abg_sheardriven_beta} and \ref{eq:omega_sheardriven}. However, strong alignment of ${\bm{\tilde{\omega}}}$ with ${\tilde{\bm{\beta}}}$ also occurs in isotropic turbulence for the unfiltered gradients \citep{meneveau_lagrangian_2011,Xu2011}. Therefore, the results in  Fig. \ref{fig:Ealign_omega} (a) for $\ell/\eta = 0$ do not necessarily confirm the prediction of the model of \cite{Okino2019}. However, in isotropic turbulence the alignments between $\tilde{\bm{\omega}}$ and ${\tilde{\bm{\beta}}}$ weaken as $\ell/\eta$ is increased \citep{Danish2018}. By contrast, Fig. \ref{fig:Ealign_omega} (b) shows that the alignment between $\tilde{\bm{\omega}}$ and ${\tilde{\bm{\beta}}}$ strengthens slightly in going from $\ell/\eta=0$ to $\ell/\eta\approx 90$. Hence the strong preferential alignment of $\tilde{\bm{\omega}}$ with ${\tilde{\bm{\beta}}}$ at larger scales does seem to be explained by strong layering in the flow due to stratification, as captured by the model of \cite{Okino2019}.

\subsection{Energetic mechanisms conditioned on alignment}

\begin{figure}
\flushright
\includegraphics[trim = {15cm 6.5cm 2cm 0cm}, scale=0.4,clip]{Legends/LegendA.pdf}
\put(-325,18){{\footnotesize $\langle X \rangle_\xi \mathcal{P}(\xi;\ell)/\langle \epsilon \rangle$}}
\put(-325,8){{\footnotesize $\langle X \rangle_\xi/\langle \epsilon \rangle$}}
\put(-235,18){{\footnotesize $Pr_1$}}
\put(-235,8){{\footnotesize $Pr_1$}}
\put(-160,18){{\footnotesize $Pr_2$}}
\put(-160,8){{\footnotesize $Pr_2$}}
\put(-80,18){{\footnotesize $Pr_3$}}
\put(-80,8){{\footnotesize $Pr_3$}}

\includegraphics[trim = {1cm 0.5cm 0.4cm 1.5cm}, scale=0.65,clip]{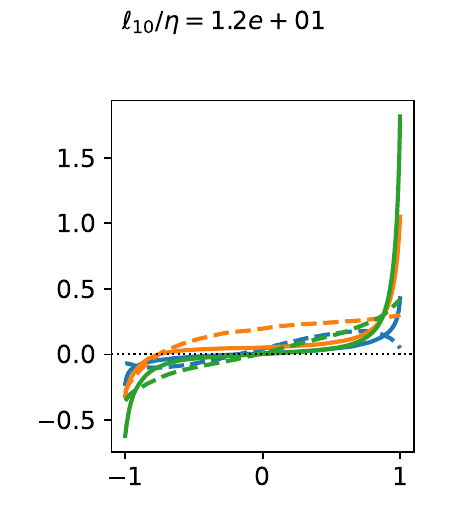}
\put(-135,50){\rotatebox{90}{{\footnotesize $X = \mathcal{B}^{sf}$}}}
\put(-125,125){{\footnotesize (a)}}
\put(-60,-5){\footnotesize $\xi$}
\includegraphics[trim = {0.4cm 0.5cm 0.5cm 1.5cm}, scale=0.65,clip]{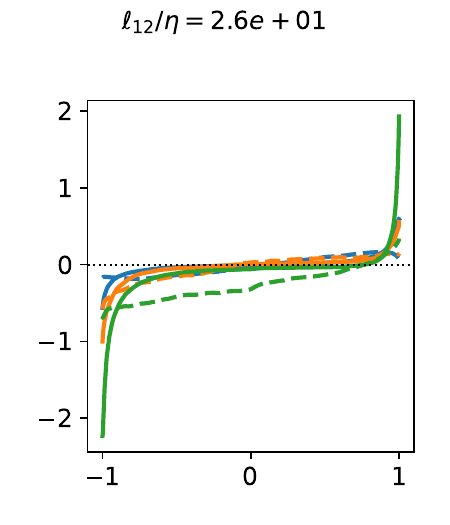}
\put(-125,125){{\footnotesize (b)}}
\put(-62,-5){\footnotesize $\xi$}
\includegraphics[trim = {0.4cm 0.5cm 0.5cm 1.5cm}, scale=0.65,clip]{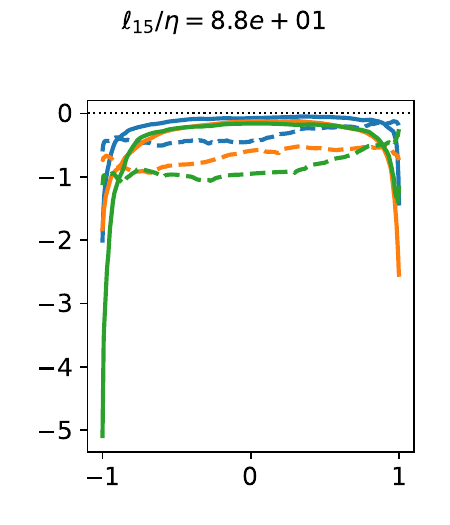}
\put(-125,125){{\footnotesize (c)}}
\put(-64,-5){\footnotesize $\xi$}
\captionsetup{width=\linewidth} 
\caption{PDF-weighted and unweighted averages conditioned on $\xi={\bm{e}}_{\tilde{\bB}} \bm{\cdot} {\bm{e}}_{z}$ of the sub-filter scale buoyancy flux $ \mathcal{B}^{sf}$ as a function of $\xi$.
Panels (a), (b) and (c) correspond to filtering at scales $\ell /\eta \approx 10$,  $\ell /\eta \approx 30$, and $\ell /\eta \approx 90$, respectively.
Data corresponds to $Pr_1=1$ (blue), $Pr_2=7$ (orange), $Pr_3=50$ (green).}
\label{fig:BsgsBfs_tot}
\end{figure}

Figure \ref{fig:BsgsBfs_tot} shows the results for $\langle \mathcal{B}^{sf}\rangle_\xi \mathcal{P}(\xi;\ell)$ and $\langle \mathcal{B}^{sf}\rangle_\xi$  as a function of $\xi$, and for scales $\ell /\eta \approx 10$ (a), $\ell /\eta \approx 30$ (b), and $\ell /\eta \approx 90$ (c). For $\ell /\eta \approx 10$ and $\ell /\eta \approx 30$, regions with strong alignment $\xi \approx +1$ correspond to regions with a strong reverse buoyancy flux that converts SFS APE to vertical TKE. \cite{Bhattacharjee_2026a} analysed the mechanism associated with this in detail and their results showed that the magnitude of the reverse flux increases in going from $Pr=1$ to $Pr=7$. Our results show that this continues to increase in going from $Pr=7$ to $Pr=50$. Indeed, for $Pr=50$, the reverse flux events near $\xi\approx +1$ are very strong, with values almost twice $\langle \epsilon \rangle$. The forward buoyancy flux in regions of strong misalignment $\xi\approx -1$ also monotonically increases in magnitude when going from $Pr=1$ to $Pr=50$. At the larger scale $\ell /\eta \approx 90$ the reverse buoyancy flux vanishes and the conditional averages are negative at all $\xi$. This is associated with the fact that at larger scales the buoyancy flux is the dominant source term in the SFS APE equation. See \cite{Bhattacharjee_2026a} for a detailed discussion of these terms. It is tempting to think that the strong reverse flux in regions where $\langle \mathcal{B}^{sf}\rangle_\xi>0$ at $\xi \approx +1$ suggests that regions of strong alignment must occur in regions where the local flow is strongly unstable. However, we will show later that this is not the case, and that a non-trivial relationship exists between the local alignment and local stability.

%
\begin{figure}
\centering
\includegraphics[trim = {20cm 6.5cm 2cm 0cm}, scale=0.4,clip]{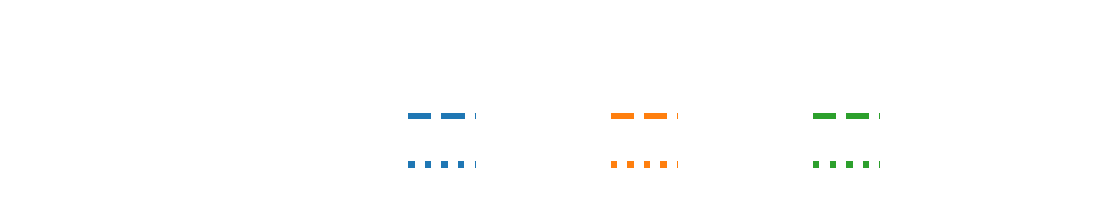}
\put(-200,18){{\footnotesize $\langle X \rangle_\xi \mathcal{P}(\xi;\ell)/\langle \epsilon \rangle$}}
\put(-200,8){{\footnotesize $\langle X \rangle_\xi/\langle \epsilon \rangle$}}
\put(-115,18){{\footnotesize $Pr_1$}}
\put(-115,8){{\footnotesize $Pr_1$}}
\put(-80,18){{\footnotesize $Pr_2$}}
\put(-80,8){{\footnotesize $Pr_2$}}
\put(-40,18){{\footnotesize $Pr_3$}}
\put(-40,8){{\footnotesize $Pr_3$}}

\includegraphics[trim = {-0.9cm 0.5cm 8.6cm 1cm}, scale=0.65,clip]{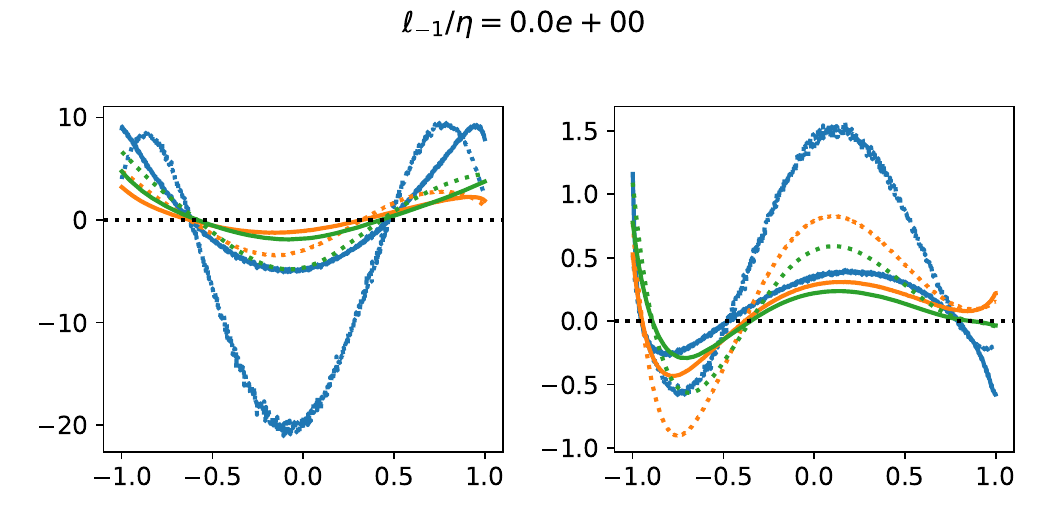}
\put(-170,115){{\footnotesize (a)}}
\put(-170,55){\rotatebox{90}{{\footnotesize $X = P^f_{z}$}}}
\hfill
\includegraphics[trim = {8.7cm 0.5cm -0.9cm 1cm}, scale=0.65,clip]{Figures_cond/Pr1750Fr1_pdf_FSppbAvv_cond_eBez999.pdf}
\put(-185,115){{\footnotesize (b)}}
\put(-190,55){\rotatebox{90}{{\footnotesize $X = P^f_{b,z}$}}}
\put(-275,-8){\footnotesize $\xi$}
\put(- 90,-8){\footnotesize $\xi$}
\captionsetup{width=\linewidth} 
\caption{Conditional average of the vertical-component of the  FS 
(a) pressure redistribution $P^f_z$, and (b) buoyancy-pressure-redistribution $P^f_{b,z}$ $\ell = 0$.
Data corresponds to $Fr\approx 0.16$ for $Pr_1=1$ (blue), $Pr_2=7$ (orange), $Pr_3=50$ (green).}
\label{fig:pAvv_cond}
\end{figure}


The data for the SFS pressure redistribution terms $\langle P^{sf}_z\rangle_\xi \mathcal{P}(\xi;\ell)$ and $\langle P^{sf}_z\rangle_\xi$ are quite noisy, and are therefore not shown. 
The lack of statistical convergence is mainly due to the fact that the main contribution to this pressure term comes from large-scales, as shown earlier. Therefore, as an alternative, in Figure \ref{fig:pAvv_cond} (a) we show results for the FS pressure terms $\langle P^{f}_z\rangle_\xi \mathcal{P}(\xi;\ell)$ and $\langle P^{f}_z\rangle_\xi$, respectively, for $\ell/\eta=0$. Corresponding results for the horizontal term $P^{sf}_h$ are not shown since, due to incompressibility, $P^{sf}_h=-P^{sf}_z$. For the unconditioned average, $\langle P^{f}_z\rangle>0$, providing the mechanism that transfers kinetic energy from the horizontal to the vertical velocity field. The results for $\langle P^{f}_z\rangle_\xi$ show that the strongest redistribution events occur in regions with weak alignment $|\xi|\ll 1$. In these regions $\langle P^{f}_z\rangle_\xi<0$ such that kinetic energy is transferred from the vertical back to the horizontal velocity field, while regions where $|\xi|=O(1)$ are regions where $\langle P^{f}_z\rangle_\xi>0$. Although the magnitude of $\langle P^{f}_z\rangle_\xi$ is much stronger for $|\xi|\ll 1$ than for $|\xi|=O(1)$, the contribution from the former dominates and ensures that $\langle P^{f}_z\rangle>0$ due to the probability of $|\xi|=O(1)$ being much greater than that of $|\xi|\ll 1$, as shown earlier. 

The pressure field in stratified turbulence involves separate contributions from the nonlinear acceleration term and the buoyancy terms in the momentum equation \citep{YKBZ2025,YKBZ2026}. In contrast to the nonlinear component which always drives the flow towards isotropy, \cite{YKBZ2025} observed that the buoyancy component of the pressure-strain correlation promotes large-scale anisotropy under conditions of strong stratification, but isotropy at weaker stratification. It is of interest to understand how this contribution depends on the direction of the gradient of the fluctuating density field. Therefore, in Figure \ref{fig:pAvv_cond} (b) we show $\langle P^{f}_{b,z}\rangle_\xi$, where $P^{f}_{b,z}$ denotes the contribution to the pressure redistribution term that contains the buoyancy contribution to the pressure field but not the nonlinear contribution. Comparing the results to those in plot (a) reveals that the strong negative values for $\langle P^{f}_z\rangle_\xi$ at $|\xi|\ll 1$ arise due to the nonlinear contribution to the pressure field, while the buoyancy contribution at $|\xi|\ll 1$ leads to the more expected behaviour of kinetic energy being transferred from the vertical to horizontal fields. 

The results in Figure \ref{fig:pAvv_cond} (a) also show that while the dependence of $\langle P^{f}_z\rangle_\xi$ on $\xi$ is qualitatively similar for each $Pr$, the dependence of $\langle P^{f}_z\rangle_\xi$ on $\xi$ is weaker. This is due on the one hand to the pressure contribution coming mainly from larger scales in the flow, while $\tilde{\bB}$ is dominated by the smallest scales which become smaller as $Pr$ is increased. The increased separation between the scales that dominate $P^{f}_z$ and $\tilde{\bB}$ as $Pr$ increases therefore leads to a decreased dependence of $\langle P^{f}_z\rangle_\xi$ on $\xi$ as $Pr$ increases. On the other hand, it is also due to the increasing uniformity of the distribution of $\xi$ for $\ell/\eta=0$ as $Pr$ increases.

\begin{figure}
\flushright
\includegraphics[trim = {15cm 6.5cm 2cm 0cm}, scale=0.4,clip]{Legends/LegendA.pdf}
\put(-325,18){{\footnotesize $\langle X \rangle_\xi \mathcal{P}(\xi;\ell)/\langle \epsilon \rangle$}}
\put(-325,8){{\footnotesize $\langle X \rangle_\xi/\langle \epsilon \rangle$}}
\put(-235,18){{\footnotesize $Pr_1$}}
\put(-235,8){{\footnotesize $Pr_1$}}
\put(-160,18){{\footnotesize $Pr_2$}}
\put(-160,8){{\footnotesize $Pr_2$}}
\put(-80,18){{\footnotesize $Pr_3$}}
\put(-80,8){{\footnotesize $Pr_3$}}

\flushleft
\includegraphics[trim = {0mm 0mm 0mm 15mm}, scale=0.605,clip]{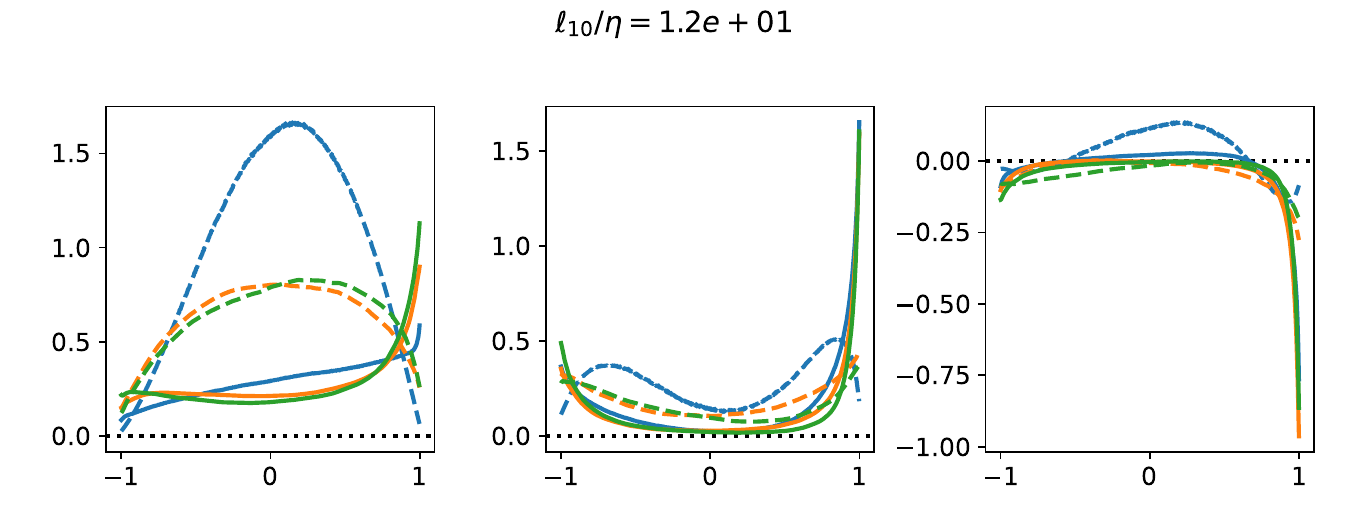}
\put(-395,65){\rotatebox{90}{\footnotesize $X_\xi$}}
\put(-390,120){{\footnotesize (a)}}
\put(-362,85){\textcolor{gray}{{\footnotesize $X=\Pi_h$}}}
\put(-260,120){{\footnotesize (b)}}
\put(-230,85){\textcolor{gray}{{\footnotesize $X=\Pi_{z}$}}}
\put(-130,120){{\footnotesize (c)}}
\put(-105,85){\textcolor{gray}{{\footnotesize $X=\Pi_{r,z}$}}}\\
\includegraphics[trim = {0mm 0mm 0mm 15mm}, scale=0.605,clip]{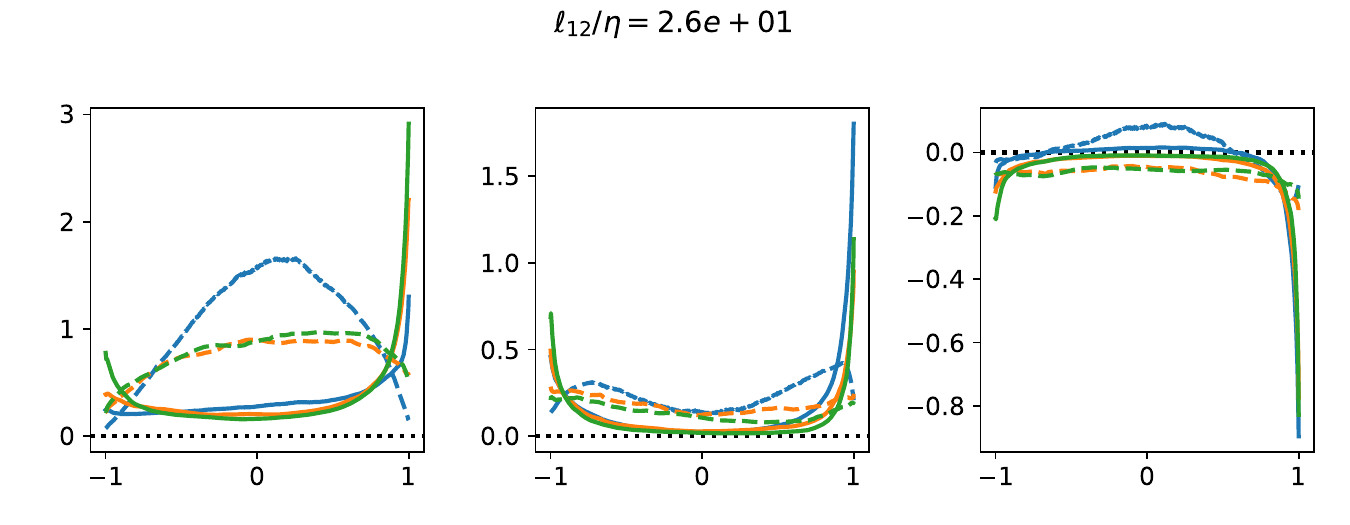}
\put(-395,65){\rotatebox{90}{\footnotesize $X_\xi$}}
\put(-390,120){{\footnotesize (d)}}
\put(-315,-2){\footnotesize $\xi$}
\put(-362,85){\textcolor{gray}{{\footnotesize $X=\Pi_h$}}}
\put(-260,120){{\footnotesize (e)}}
\put(-190,-2){\footnotesize $\xi$}
\put(-230,85){\textcolor{gray}{{\footnotesize $X=\Pi_{z}$}}}
\put(-130,120){{\footnotesize (f)}}
\put(-60,-2){\footnotesize $\xi$}
\put(-105,85){\textcolor{gray}{{\footnotesize $X=\Pi_{r,z}$}}}\\
\captionsetup{width=\linewidth} 
\caption{Plots of scale-to-scale flux for (a,d) the horizontal TKE $\Pi_h$, (b,e) the vertical TKE $\Pi_z$, and (c,f) the vertical component of the flux-redistribution $\Pi_{r,z}$, versus $\xi={\bm{e}}_{\tilde{\bB}} \bm{\cdot} {\bm{e}}_{z}$ for (a-c) $\ell/\eta \approx 10$ and (d-f) $\ell/\eta \approx 30$. 
Blue, orange and green curves correspond to $Pr_1=1$, $Pr_2=7$, and $Pr_3=50$, respectively.
Dashed curves correspond to the conditional expectation $\langle X \rangle_\xi$, whereas solid curves correspond to the PDF-weighted average $\langle X \rangle_\xi \mathcal{P}(\xi;\ell)$, for $X=\Pi_h,\Pi_{r,z},\Pi_z$.
\label{fig:Pihrv_xi} }
\end{figure}

Figure \ref{fig:Pihrv_xi} (a), (d) show results for $\langle \Pi_h\rangle_\xi \mathcal{P}(\xi;\ell)$ and $\langle \Pi_h\rangle_\xi$, with the other subplots showing the corresponding results for the fluxes $\Pi_z$ and $\Pi_{r,z}$. Plots (a)-(c) are for $\ell/\eta\approx 10$ while plots (d)-(f) are for $\ell/\eta\approx 30$, corresponding to scales below and above the Ozmidov scale, respectively(the results for $\ell/\eta\approx 90$ are too noisy to consider). If the density field were passive, and the flow were isotropic at scales $\ell<l_O$ then we might expect to see $\langle \Pi_h\rangle_\xi$ being almost independent of $\xi$. The results for $\ell/\eta\approx 10$ however show that this is not the case. To some extent this is not surprising since we have already seen evidence that the small-scales of the flow are not isotropic. What is striking, however, is that $\langle \Pi_h\rangle_\xi$ drops to almost zero for $|\xi|\approx 1$ for $Pr=1$. For $Pr=7,50$,  $\langle \Pi_h\rangle_\xi$ drops to small values for $\xi\approx -1$, but larger values for $\xi\approx +1$. On the other hand, the results for $\langle \Pi_z\rangle_\xi$ show the opposite behaviour, with $\langle \Pi_z\rangle_\xi$ minimum for $\xi\approx 0$, and largest for $|\xi|\approx 1$ (although this dependence on $\xi$ weakens for $Pr=7,50$ and at $\ell/\eta\approx 30$). This shows that in stratified turbulence the local alignment ${\bm{e}}_{\tilde{B}} \bm{\cdot} \ez$ plays an important role in controlling the local TKE interscale flux. Moreover, the preference for states where $|{\bm{e}}_{\tilde{B}} \bm{\cdot} \ez|\approx 1$ shown earlier means that the formation of ramp-cliff structures in the flow causes the average energy flux $\langle \Pi_h\rangle$ to be much weaker than it would otherwise be (especially for $Pr=1$), because states where $|{\bm{e}}_{\tilde{B}} \bm{\cdot} \ez|\approx 1$ also correspond to states where $\langle \Pi_h\rangle_\xi$ is small. This is analagous to the known fact that the energy cascade in isotropic turbulence is inefficient \citep{Ballouz_Ouellette_2018}, associated in part with the preference for the vorticity to align with the intermediate strain-rate eigendirection rather than the extensional direction which has the largest positive eigenvalue.

The results in Figure \ref{fig:Pihrv_xi} c,f show that for $Pr=1$, $\langle \Pi_{r,z}\rangle$ is negative for $|\xi|\approx 1$ (corresponding to an upscale vertical TKE flux), but positive for $|\xi|\ll1 $  (corresponding to a downscale vertical TKE flux). For $Pr=7,50$, however, it becomes negative at all $\xi$, meaning that it always acts to send vertical TKE upscale. The most striking feature, however, is that $\langle \Pi_{r,z}\rangle\mathcal{P}(\xi;\ell)=O(-\langle\epsilon\rangle)$ for $\xi\approx +1$, corresponding to a strong upscale vertical TKE flux, and since $\Pi_{r,h}=-\Pi_{r,z}$, a strong downscale horizontal TKE flux. This contribution has a strong cancellation effect on $\langle \Pi_{z}\rangle\mathcal{P}(\xi;\ell)$ for $\xi\approx +1$, such that the total vertical TKE flux due to nonlinearity $(\langle \Pi_{z}\rangle+\langle \Pi_{r,z}\rangle)\mathcal{P}(\xi;\ell)$ is quite small for $\xi\approx +1$. On the other hand, $\langle \Pi_{r,h}\rangle\mathcal{P}(\xi;\ell)$ strongly contributes to the total horizontal TKE flux due to nonlinearity $(\langle \Pi_{h}\rangle+\langle \Pi_{r,h}\rangle)\mathcal{P}(\xi;\ell)$ for $\xi\approx +1$.

\begin{figure}
\centering
\includegraphics[trim = {5cm 7cm 2cm 0cm}, scale=0.4,clip]{Legends/LegendA.pdf}
\put(-325,12){{\footnotesize $\langle X \rangle_\xi \mathcal{P}(\xi;\ell)/\langle \chi \rangle$}}
\put(-325,2){{\footnotesize $\langle X \rangle_\xi/\langle \chi \rangle$}}
\put(-235,12){{\footnotesize $Pr_1$}}
\put(-235,2){{\footnotesize $Pr_1$}}
\put(-160,12){{\footnotesize $Pr_2$}}
\put(-160,2){{\footnotesize $Pr_2$}}
\put(-80,12){{\footnotesize $Pr_3$}}
\put(-80,2){{\footnotesize $Pr_3$}}

\flushleft
\includegraphics[trim = {0.2cm 0cm 0cm 1.2cm}, scale=0.6,clip]{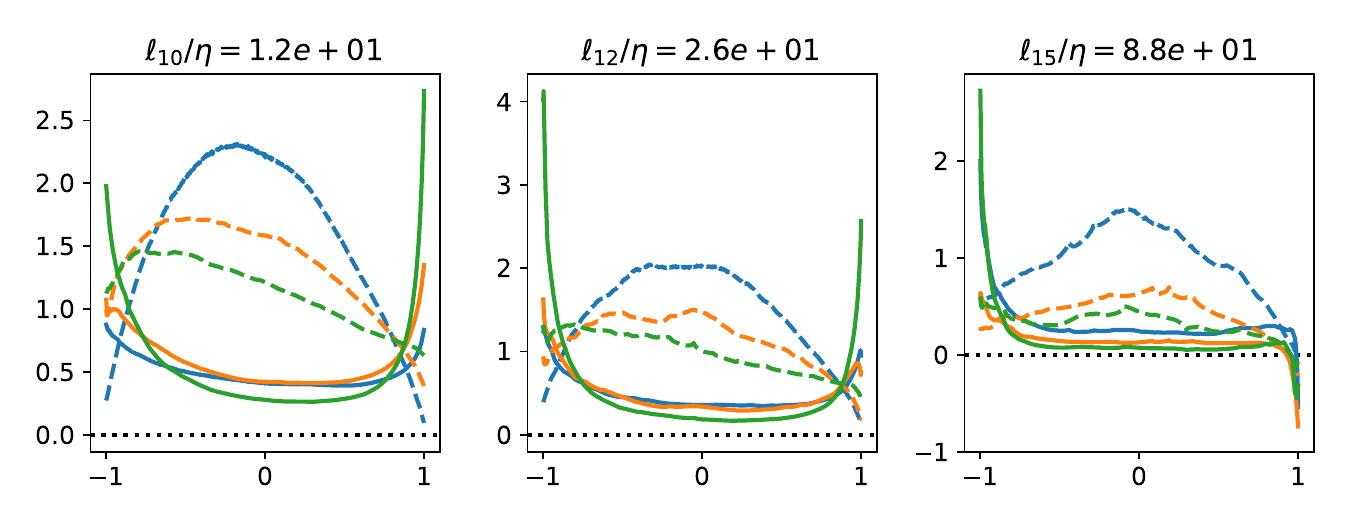}
\put(-390,60){\rotatebox{90}{\footnotesize $X = \Pi_\phi$}}
\put(-385,130){{\footnotesize (a)}}
\put(-255,130){{\footnotesize (b)}}
\put(-125,130){{\footnotesize (c)}}
\put(-360,120){\textcolor{gray}{{\footnotesize $\ell/\eta \approx 10$}}}
\put(-230,120){\textcolor{gray}{{\footnotesize $\ell/\eta \approx 30$}}}
\put(-100,120){\textcolor{gray}{{\footnotesize $\ell/\eta \approx 90$}}}
\put(-315,0){\footnotesize $\xi$}
\put(-190,0){\footnotesize $\xi$}
\put(- 62,0){\footnotesize $\xi$}
\captionsetup{width=\linewidth} 
\caption{Plots of scale-to-scale flux for APE, $\Pi_\phi$, normalized by the APE destruction rate $\langle \chi \rangle$ versus $\xi={\bm{e}}_{\tilde{\bB}} \bm{\cdot} {\bm{e}}_{z}$ for (a) $\ell/\eta \approx 10$, (b) $\ell/\eta \approx 30$, and (c) $\ell/\eta \approx 90$. 
Blue, orange and green curves correspond to $Pr_1=1$, $Pr_2=7$, and $Pr_3=50$, respectively.
Dashed curves correspond to the conditional expectation $\langle \Pi_\phi \rangle_\xi$, whereas solid curves correspond to the PDF-weighted average $\langle \Pi_\phi \rangle_\xi \mathcal{P}(\xi;\ell)$.
}
\label{fig:Piphi_cond}
\end{figure}

Figure \ref{fig:Piphi_cond} shows results for $\langle \Pi_\phi\rangle_\xi \mathcal{P}(\xi;\ell)$ and $\langle \Pi_\phi\rangle_\xi$ at scales $\ell/\eta\approx 10$ (a), $\ell/\eta\approx 30$ (b), and $\ell/\eta\approx 90$ (c). The results for $Pr=1$ and for $\ell/\eta\approx 10$ and $\ell/\eta\approx 30$ show that $\langle \Pi_\phi\rangle_\xi$ is maximum for $|\xi|\ll 1$, and approaches small values for $|\xi|\to 1$. As $Pr$ increases, the maximum shifts towards more negative values of $\xi$. The growth of the values for $\xi\to -1$ can be understood by noting that in this regime of $\xi$ the buoyancy flux $\langle \mathcal{B}^{sf} \rangle_\xi$ is negative and grows in magnitude with increasing $Pr$, corresponding to an increased amount of TKE being transferred to APE as $Pr$ is increased, allowing more APE to be passed downscale through $\Pi_\phi$. Furthermore, just as for the horizontal TKE flux, the preference for states where $|{\bm{e}}_{\tilde{B}} \bm{\cdot} \ez|\approx 1$ shown earlier means that the formation of ramp-cliff structures in the flow causes the average energy flux $\langle \Pi_\phi\rangle$ to be much weaker than it would otherwise be, because states where $|{\bm{e}}_{\tilde{B}} \bm{\cdot} \ez|\approx 1$ also correspond to states where $\langle \Pi_\phi\rangle_\xi$ is relatively small.

At the larger scale $\ell/\eta\approx 90$, Figure~\ref{fig:Piphi_cond} (c) shows that an inverse flux emerges for $\xi\approx +1$. Although the magnitude of $\langle \Pi_\phi \rangle_\xi$ is relatively small in these regions ($\ll \langle\chi\rangle$), because the probability of regions where $\xi\approx +1$ is high, the weighted contribution is significant, with $|\langle \Pi_\phi \rangle_\xi \mathcal{P}(\xi;\ell)|=O(\langle\chi\rangle)$ for $\xi\approx +1$. Clearly the local alignment ${\bm{e}}_{\tilde{B}} \bm{\cdot} \ez$ plays a very important role at larger scales on the APE in the flow, with $\xi\approx -1$ and $\xi\approx +1$ distinguishing regions of strong downscale and upsale APE flux, respectively.

\begin{figure}
\centering
\includegraphics[trim = {15cm 6.5cm 2cm 0cm}, scale=0.4,clip]{Legends/LegendA.pdf}
\put(-252,25){{\footnotesize $Pr_1$}}
\put(-235,18){{\footnotesize $\langle X \rangle_\xi \mathcal{P}(\xi;\ell)$}}
\put(-235,8){{\footnotesize $\langle X \rangle_\xi$}}
\put(-177,25){{\footnotesize $Pr_2$}}
\put(-160,18){{\footnotesize $\langle X \rangle_\xi \mathcal{P}(\xi;\ell)$}}
\put(-160,8){{\footnotesize $\langle X \rangle_\xi$}}
\put(-97,25){{\footnotesize $Pr_3$}}
\put(-80,18){{\footnotesize $\langle X \rangle_\xi \mathcal{P}(\xi;\ell) $}}
\put(-80,8){{\footnotesize $\langle X \rangle_\xi$}}

\flushleft
\includegraphics[trim = {0mm 7.7cm 0cm 2.15cm}, scale=0.61,clip]{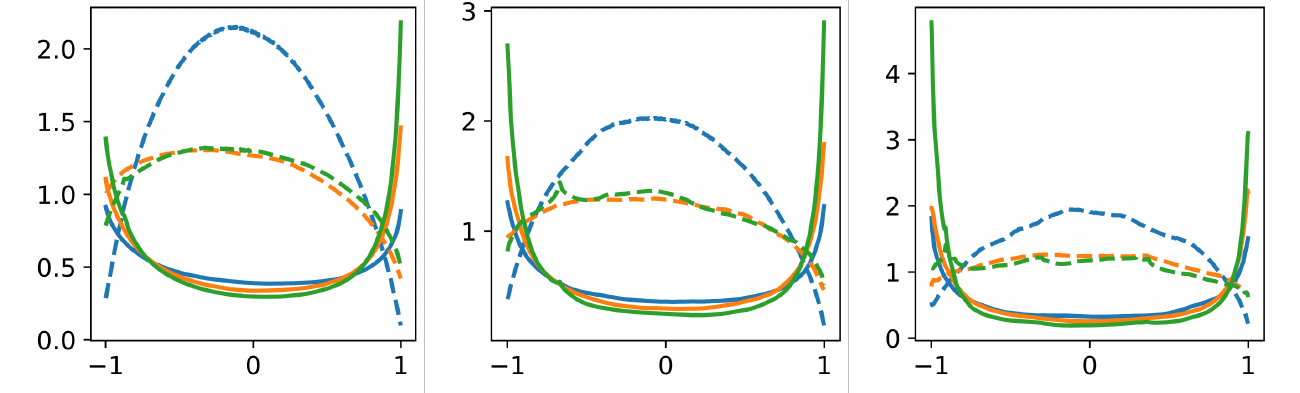}
\put(-382,45){\rotatebox{90}{{\footnotesize $\langle \varepsilon_\phi \rangle_\xi / \langle {\chi} \rangle$}}}
\put(-380,110){{\footnotesize (a)}}
\put(-258,110){{\footnotesize (b)}}
\put(-130,110){{\footnotesize (c)}}

\includegraphics[trim = {0mm 0cm 0cm 9.5cm}, scale=0.61,clip]{Figures_cond/Pr1750Fr1_PDF_SFS_chieps_vs_eBez101215.pdf}
\put(-380,45){\rotatebox{90}{{\footnotesize $\langle \varepsilon_K \rangle_\xi / \langle {\epsilon} \rangle$}}}
\put(-380,117){{\footnotesize (d)}}
\put(-258,117){{\footnotesize (e)}}
\put(-130,117){{\footnotesize (f)}}
\put(-325,110){\textcolor{gray}{{\footnotesize $\ell/\eta \approx 10$}}}
\put(-205,110){\textcolor{gray}{{\footnotesize $\ell/\eta \approx 30$}}}
\put(- 80,110){\textcolor{gray}{{\footnotesize $\ell/\eta \approx 90$}}}
%
%
\put(-310,0){\footnotesize $\xi$}
\put(-188,0){\footnotesize $\xi$}
\put(- 64,0){\footnotesize $\xi$}
\captionsetup{width=\linewidth} 
\caption{Plots of the SFS APE dissipation rate $ {\varepsilon_\phi}$ (top row) and the SFS TKE dissipation rate  ${\varepsilon_K}$ (bottom row), conditioned on the alignment of the density gradient with the vertical $\xi={\bm{e}}_{\tilde{\bB}} \bm{\cdot} {\bm{e}}_{z}$ for $Pr_1=1$ (blue), $Pr_2=7$ (orange), and $Pr_3=50$ (green), for $Fr\approx 0.16$.
Panels (a), (d) correspond to filtering at scales $\ell /\eta \approx 10$; (b), (e) correspond to filtering at scales $\ell /\eta \approx 30$; (c), (f) correspond to filtering at scales $\ell /\eta \approx 90$, respectively. 
Solid curves correspond to the PDF-weighted average $\langle X \rangle_\xi \mathcal{P}(\xi;\ell)$, whereas dashed curves correspond to the conditional expectation $\langle X \rangle_\xi$, for $X=\varepsilon_\phi,\varepsilon_K$.
}
\label{fig:ChiEps_contr_cond}
\end{figure}

Figure \ref{fig:ChiEps_contr_cond} shows the results for the weighted and unweighted conditional averages of the SFS APE $\varepsilon_\phi$ (top row) and SFS TKE $\varepsilon_K$ (bottom row) dissipation rates. The dependence of the unweighted averages on $\xi$ is similar to that of the energy fluxes,
with $\langle\varepsilon_K\rangle_\xi$ and $\langle\varepsilon_\phi\rangle_\xi$ both being maximum for $\xi\ll1$, and taking on small values for $|\xi|\to 1$. This suggests that the strongest SFS velocity and fluctuating buoyancy gradients occur in the regime $\xi\ll1$. The dependence on $\xi$ weakens, however, as $Pr$ is increased, with stronger dissipation taking place in regions of strong alignment $|\xi|\to 1$. This is likely associated with the earlier observation that the PDF $\mathcal{P}(\xi;\ell)$ becomes more uniform with increasing $Pr$, so that regions with the strongest dissipation become less dependent on the local alignment indicated by $\xi$. At scales where the dissipation rates are balanced by the energy fluxes, the behaviour of $\langle\varepsilon_K\rangle_\xi$ and $\langle\varepsilon_\phi\rangle_\xi$ make sense since the dissipation rates will be largest in regions of $\xi$ where the energy fluxes are largest. The weighted averages show that at $\ell/\eta\approx 10$, most of the contribution to $\langle\varepsilon_K\rangle$ comes from regions with $\xi\approx +1$, where the small-scale reverse buoyancy flux is producing vertical TKE. At the larger scale $\ell/\eta\approx 90$, regions with $\xi\approx -1$ and $\xi\approx +1$ contribute more evenly to $\langle\varepsilon_K\rangle$ as $Pr$ increases. This is quite surprising because it means that a significant contribution to $\langle\varepsilon_K\rangle$ at larger scales comes from regions where $\langle \mathcal{B}^{sf} \rangle_\xi$ is strongly negative. However, noting again that $\langle\varepsilon_K\rangle_\xi$ is almost uniform at $\ell/\eta\approx 90$ for the higher $Pr$ cases, we see that the strong contribution to $\langle\varepsilon_K\rangle$ from regions where $\xi\approx -1$ is simply because of the relatively high probability of $\xi\approx -1$ regions. The strongest contributions to $\langle\varepsilon_\phi\rangle$ come from both $\xi\approx -1$  and $\xi\approx +1$ regions. This makes sense because in both of these regions $\langle \mathcal{B}^{sf} \rangle_\xi$ is strongly negative, such that there is a strong source of APE that can be sent downscale and dissipated. 

\subsection{Mixing efficiency conditioned on alignment}

\begin{figure}
\centering
\includegraphics[trim = {8cm 7.5cm 2cm 1cm}, scale=0.4,clip]{Legends/LegendA.pdf}
\put(-235,5){{\footnotesize $Pr_1$}}
\put(-160,5){{\footnotesize $Pr_2$}}
\put(- 80,5){{\footnotesize $Pr_3$}}

\includegraphics[trim = {0mm 0mm 0mm 11mm}, scale=0.605,clip]{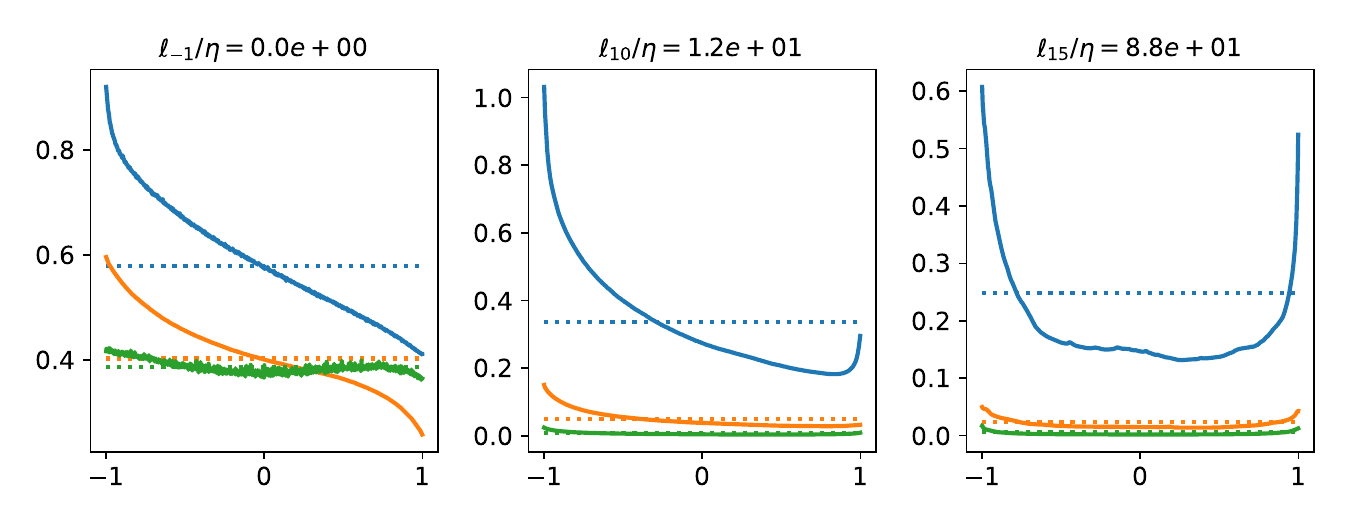}
\put(-395,65){\rotatebox{90}{\footnotesize $\Gamma_\xi$}}
\put(-390,135){{\footnotesize (a)}}
\put(-317,-2){\footnotesize $\xi$}
\put(-332,120){\textcolor{gray}{{\footnotesize $\ell/\eta = 0$}}}
\put(-260,135){{\footnotesize (b)}}
\put(-190,-2){\footnotesize $\xi$}
\put(-220,120){\textcolor{gray}{{\footnotesize $\ell/\eta \approx 10$}}}
\put(-130,135){{\footnotesize (c)}}
\put(-60,-2){\footnotesize $\xi$}
\put(- 85,120){\textcolor{gray}{{\footnotesize $\ell/\eta \approx 90$}}}\\
\captionsetup{width=\linewidth} 
\caption{Plots of the average FS mixing coefficient $\tilde{\Gamma}_\xi =\langle \tilde{\chi} \rangle_\xi / \langle \tilde{\epsilon} \rangle_\xi$ conditioned on the alignment of the fluctuating density gradient $\xi={\bm{e}}_{\tilde{\bB}} \bm{\cdot} {\bm{e}}_{z}$ (solid lines) filtered at scales 
(a) $\ell /\eta = 0$, (b) $\ell /\eta \approx 10$, and (c) $\ell /\eta \approx 90$,
for $Pr_1=1$ (blue), $Pr_2=7$ (orange), and $Pr_3=50$ (green), for $Fr\approx 0.16$.
Dotted lines show the ensemble-averaged value of the mixing coefficient $\Gamma$.
}
\label{fig:GammaFS_cond}
\end{figure}

The mixing efficiency is an important quantity in stably stratified turbulence, quantifying how much of the energy supplied to the flow is dissipated by the APE field through irreversible mixing (which corresponds to the rate at which irreversible mixing acts to increase the background potential energy \citep{Winters_1995}) compared with the TKE dissipated due to viscous stresses. To analyse how the mixing efficiency depends on the alignment through $\xi$ as well as on scale, in Figure \ref{fig:GammaFS_cond} we show results for $\Gamma_\xi\equiv\langle \widetilde{\chi}\rangle_\xi/\langle  \widetilde{\epsilon}\rangle_\xi$ for filter scales (a) $\ell /\eta = 0$, (b) $\ell /\eta \approx 10$, and (c) $\ell /\eta \approx 90$. 
For $Pr=1$ and $\ell/\eta=0$ (unfiltered case), the results indicate that irreversible mixing is strongest in regions where $\xi \approx -1 $. However, as $Pr$ increases, $\Gamma_\xi$ becomes increasingly uniform, and is almost independent of $\xi$ for $Pr=50$. This is due to the PDF of $\xi$ becoming increasingly uniform for $\ell/\eta=0$ as $Pr$ increases, as shown earlier, indicating that at the smallest scales in the flow, the ramp-cliff asymmetry vanishes in the limit $Pr\to\infty$. As $\ell/\eta$ is increased, $\Gamma_\xi$ reduces dramatically for $Pr=7,50$, while it reduces more slowly with increasing $\ell/\eta$ for $Pr=1$. This is due to the fact that as $Pr$ increases, mixing takes place at smaller and smaller scales, with the mean-field estimate for the scale at which mixing takes place being given by the Batchelor length-scale $\eta_B=Pr^{-1/2}\eta$. It is interesting that $\Gamma_\xi$ remains so large even at $\ell/\eta=90$ for $Pr=1$, especially for $\xi\approx-1$, given that this scale is much larger than the scale at which dissipation is expected to be strongest. It is also very interesting that as $\ell/\eta$ increases, $\Gamma_\xi$ actually increases at $\xi\approx +1$. The same also occurs for $Pr=7,50$, however, the increase near $\xi\approx +1$ is small for these cases. An explanation for this will be given momentarily.

\subsection{Relationship between alignment and local stability \label{sec:Res_align_and_stab}}

The question analysed in this paper is the impact that the local alignment of the fluctuating density gradient with the mean density gradient has on the energetics of stably stratified turbulence. An important consideration is how the alignment relates to both the ramp-cliff structures in the density field, and notions of local stability of the flow associated with the properties of the local vertical density gradient \citep{Petropoulos2024}. For this, we focus on the relationship between the alignments and the filtered vertical density gradients.

The statistical manifestation of ramps-cliff structures in the density field of a stably stratified turbulent flow (where the mean vertical buoyancy gradient is negative) is that the PDF of $\tilde{B}_z\equiv \ez\bm{\cdot} \widetilde{\bm B}$ is negatively skewed. Regions where $\tilde{B}_z/\sigma_{\tilde{B}_z}\ll -1$ (where $\sigma_{\tilde{B}_z}\equiv \sqrt{\langle \tilde{B}_z^2\rangle}$) denote steep cliffs in the density field, while regions where $0<\tilde{B}_z/\sigma_{\tilde{B}_z}\ll 1$ are associated with well-mixed ramps. Regions where $\tilde{B}_z/\sigma_{\tilde{B}_z}\geq O(1)$ are ramps with moderate to large gradients, but the largest gradients in such ramp regions are smaller than those in cliff regions due to the asymmetry $\langle\tilde{B}_z\rangle<0$. In general, while ramp-cliffs are responsible for the asymmetric PDF of $\ez\bm{\cdot} \bm{e}_{\tilde{\bm B}}$ \citep{Bragg_2024a}, there is not a simple relationship between ramp-cliffs and the behaviour of $\ez\bm{\cdot} \bm{e}_{\tilde{\bm B}}$ because ramp-cliffs are associated with the magnitude of $\widetilde{\bm B}$ and not just its direction. 

\begin{figure}
\centering
\includegraphics[trim = {15cm 6.5cm 2cm 0cm}, scale=0.4,clip]{Legends/LegendA.pdf}
\put(-235,18){{\footnotesize $Pr_1, \xi \in X_w$}}
\put(-235,8){{\footnotesize $Pr_1, \xi \in X_s$}}
\put(-160,18){{\footnotesize $Pr_2, \xi \in X_w$}}
\put(-160,8){{\footnotesize $Pr_2, \xi \in X_s$}}
\put(-80,18){{\footnotesize $Pr_3, \xi \in X_w$}}
\put(-80,8){{\footnotesize $Pr_3, \xi \in X_s$}}

\includegraphics[trim = {0mm 0cm 0cm 1.1cm}, scale=0.49,clip]{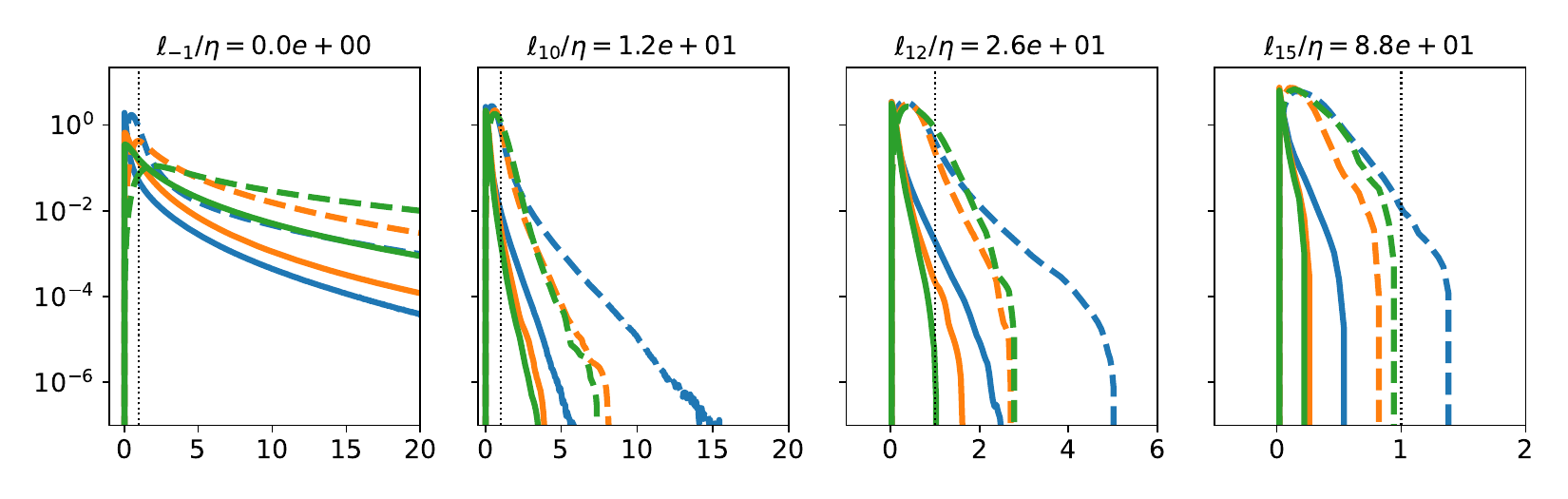} 
\put(-375,105){{\footnotesize (a)}}
\put(-283,105){{\footnotesize (b)}}
\put(-191,105){{\footnotesize (c)}}
\put(-102,105){{\footnotesize (d)}}
\put(-390,45){\rotatebox{90}{{\footnotesize $\Psi_\xi(\chi;\ell)$}}}
\put(- 50,-2){{\footnotesize $\chi/N$}}
\put(-145,-2){{\footnotesize $\chi/N$}}
\put(-240,-2){{\footnotesize $\chi/N$}}
\put(-330,-2){{\footnotesize $\chi/N$}}
\captionsetup{width=\linewidth} 
\caption{Conditional probability density function (PDF) $\Psi_\xi(\chi;\ell)$ for scales (a) $\ell/\eta = 0$, (b) $\ell/\eta \approx 10$, (c) $\ell/\eta \approx 30$, (d) $\ell/\eta \approx 90$.
Solid lines correspond to conditioning $\xi \in X_w \equiv [0,0.3)$ (weak alignment), whereas dashed lines correspond to $\xi\in X_s \equiv (0.7,1]$ (strong alignment)
for $Pr_1=1$ (blue), $Pr_2=7$ (orange), and $Pr_3=50$ (green), for $Fr\approx 0.16$.
Vertical dotted line indicates $\chi=N$, such that $\chi > N$ indicates unstable regions. }
\label{fig:drdz_condxi}
\end{figure}

The total vertical rescaled-density gradient at scale $\ell$ is given by $\nabla_z\Phi=\tilde{B}_z-N$. Stable regions $\nabla_z\Phi<0$ therefore correspond to regions where $\tilde{B}_z<N$, while unstable regions $\nabla_z\Phi>0$ correspond to regions where $\tilde{B}_z>N$. Since $\ez\bm{\cdot} \bm{e}_{\tilde{\bm B}}= \tilde{B}_z/\|\widetilde{\bm B} \|$, events with $\ez\bm{\cdot} \bm{e}_{\tilde{\bm B}}<0$ necessarily occur in stable regions, while events with $\ez\bm{\cdot} \bm{e}_{\tilde{\bm B}}>0$ could occur in either stable or unstable regions. Figure \ref{fig:drdz_condxi} shows the conditional PDF $\Psi_\xi(\chi;\ell)\equiv \langle\delta(\tilde{B}_z-\chi) \rangle_\xi$ where the condition on the average is computed both for $\xi =\ez\bm{\cdot} \bm{e}_{\tilde{\bm B}}\in X_w \equiv [0,0.3)$, corresponding to weak misalignment, and $\xi=\ez\bm{\cdot} \bm{e}_{\tilde{\bm B}}\in X_s \equiv (0.7,1]$, corresponding to strong alignment. The PDF is defined only for $\chi\geq 0$ since it is only in this range that there may be a non-trivial relationship between the signs of $\ez\bm{\cdot} \bm{e}_{\tilde{\bm B}}$ and $\nabla_z\Phi$. The results are shown for scales (a) $\ell/\eta \ =0$, (b) $\ell/\eta \approx 10$, (c) $\ell/\eta  \approx 30$, and (d) $\ell/\eta \approx 90$ and confirm that there is no simple relationship between the local stability of the flow and the local alignment. At all scales in strongly unstable regions $\chi \gg N$, the PDF is significantly larger in regions of strong alignment than weak alignment. This shows that unstable events where $\tilde{B}_z\gg N$ are most frequently associated with strong alignment events. Nevertheless, the probability of events with $\tilde{B}_z\gg N$ and weak alignment is still significant, and grows as $Pr$ increases. Moreover, even for events with strong alignment, the peak of the PDF occurs in unstable regions $\chi < N$, i.e. strong alignment events most frequently occur in stable not unstable regions. This clearly shows then that the effect of alignment on the local flow energetics presented earlier cannot be understood simply through a trivial connection between the alignment and the local stability.

The results also show that the probability of events with strong alignment occurring in stable regions $\chi< N$ increases with increasing $\ell/\eta$ for each $Pr$. Since we would expect $\Gamma_\xi$ to be largest in stable regions, this could explain the earlier observation based on Figure \ref{fig:GammaFS_cond} that for $\xi\approx +1$, $\Gamma_\xi$ increases with increasing $\ell/\eta$. The fact that this increase is much larger for $Pr=1$ than for $Pr=7,50$ is likely simply due to the fact that for a fixed $\ell/\eta>1$, $\Gamma_\xi$ will reduce with increasing $Pr$ at all $\xi$ since the strongest scalar mixing takes place at smaller scales as $Pr$ increases.

These results also explain the surprising result shown earlier that $\langle \Pi_h\rangle_\xi$ becomes small not only for $\xi\approx -1$ but also $\xi\approx +1$. If $\xi\approx +1$ corresponded to regions of strong convection, as might be naively expected, then $\langle \Pi_h\rangle_\xi$ should be large there. The results in figure \ref{fig:drdz_condxi}  clarify that the reason this does not occur is because somewhat surprisingly, regions of strong alignment mainly occur in stable rather than unstable regions. Similar observations also explain the behaviour of the dissipation rates $\langle \varepsilon_K\rangle$ and $\langle \varepsilon_\phi\rangle$ in regions of strong alignment.

\section{Conclusions}\label{sec:Conclusions}

In this work, we have explored the role that alignments between the fluctuating and mean density gradients have on the scale-dependent energetics of stably stratified turbulence. Just as non-trivial alignments between the vorticity vector and the strain-rate eigenvectors have an important impact on the velocity gradient dynamics and TKE cascade in isotropic turbulence, the recent work of \citet{BraggdBKops_2024} has highlighted the importance of the alignment between the fluctuating and mean density gradient on mixing rates in stably stratified turbulence. \citet{Bhattacharjee_2026a} also showed that these alignments are connected to the reversal of the buoyancy flux at the small-scales of  stably stratified turbulence. Motivated by this, we have explored the role played by these alignments on the other terms in the scale-dependent TKE and APE energy equations. This is motivated by analytical results we present which suggest that there should be an impact on these other terms. To explore this detail, we used datasets from DNS of statistically stationary, stably stratified turbulence for $Pr=1,7,50$ with $Fr\approx 0.16$, and $Gn\approx 50$. A filtering operator was applied to the flow fields to consider the alignments and energetics at different scales. 

At the smallest scales, the direction of the filtered density gradient $\bm{e}_{\tilde{\bB}}$ shows strong preferential alignment with the vertical directions $\pm\ez$ (where $-\ez$ is the direction of the mean-density gradient) for $Pr=1$, with the strongest alignment occurring in the $+\ez$ direction. As discussed in \citet{BraggdBKops_2024}, this is associated with the formation of ramp-cliff structures in the fluctuating density field. However, this asymmetry weakens dramatically with increasing $Pr$. This is consistent with the behaviour for passive scalars for which the ramp-cliff structures become weaker as $Pr$ is increased \citep{Buaria2021a,Shete2022}. At larger scales, $\bm{e}_{\tilde{\bB}}$ also strongly aligns with $\pm\ez$, is strongest for $+\ez$, and the preferential alignment with $\pm\ez$ is in fact much stronger than it is at the smallest scales. Moreover, the alignment at larger scales is only relatively weakly dependent on $Pr$, and indicates that while ramp-cliff structures at the smallest scales vanish in the limit $Pr\to \infty$ (see \cite{Buaria2021a}), they persist at the large scales in this limit, with important implications for the dynamics and energetics of stably stratified turbulence. This persistent preferential alignment at larger scales, regardless of $Pr$, is to be expected since at the large scales it is the mean density gradient that is responsible for the production of the fluctuating density gradients, and this process naturally leads to preferential alignment. The results also show non-trivial alignments between $\bm{e}_{\tilde{\bB}}$ and the strain-rate eigenvectors as a function of scale. This is shown to be due to the emergence of strong layering in the flow field at larger scales, associated with vertical velocity gradients that are much larger than horizontal ones. While the flow is not entirely isotropic at the smaller scales, this strong layering is absent because buoyancy is sub-leading at these scales.

For the buoyancy flux conditioned on the alignment $\xi=\bm{e}_{\tilde{\bB}}\bm{\cdot}\ez$, at smaller scales we find that the flux is strongly negative for $\xi\approx -1$, while it is strongly positive for $\xi\approx +1$. The latter is associated with the reverse buoyancy flux at the small scales which grows stronger with increasing $Pr$. This was shown in \citet{Bhattacharjee_2026a} for $Pr=1,7$, and the present results show it for the larger case $Pr=50$. At larger scales, the buoyancy flux is negative for all $\xi$, and is associated with the buoyancy flux being the leading order source term in the APE equation at these scales. The horizontal TKE inter-scale flux conditioned on $\xi$ shows that this flux is largest for $|\xi|\ll1$, and becomes small for $|\xi|=O(1)$. Given the strong preference for states $|\xi|=O(1)$ associated with ramp-cliff structures, this means that there is a much larger probability of events occurring where the horizontal TKE flux is weak. In this sense, the formation of ramp-cliff structures in the density field is associated with an inefficiency of the horizontal TKE flux, analogous to the inefficiency of the TKE cascade in isotropic turbulence due to misalignment between the sub-grid stress and strain-rate tensors \citep{Ballouz_Ouellette_2018}. The behaviour for the APE inter-scale flux conditioned on $\xi$ is similar, except that at larger scales an upscale APE flux emerges in regions where $\xi\approx +1$. Interesting dependencies on $\xi$ of the pressure-redistribution, TKE and APE dissipation rates, and scale-dependent mixing coefficients were also observed, but which are not summarized here for brevity.

We finally examined the relationship between the local alignment $\xi=\bm{e}_{\tilde{\bB}}\bm{\cdot}\ez$ and the local stability of the flow as determined by the local total density gradient. While regions where the filtered fluctuating vertical (re-scaled) density gradient $\tilde{B}_z$ is negative are necessarily associated with stable regions (and also with $\xi<0$ since the signs of $\tilde{B}_z$ and $\xi$ necessarily correspond), regions where $\tilde{B}_z>0$ (and therefore $\xi>0$) could correspond to either stable or unstable regions, depending on how large $\tilde{B}_z$ is compared to the buoyancy frequency $N$ that is associated with the mean density gradient. We show that while the probability of regions where $\tilde{B}_z/N\gg1$ is much higher in regions where $\xi=O(1)$ than where $\xi\ll1$, regions with $\xi=O(1)$ most frequently coincide with regions where $\tilde{B}_z/N<1$. This means that, somewhat surprisingly, regions with strong alignment between $\bm{e}_{\tilde{\bB}}$ and $\ez$ occur most frequently in stable regions, not unstable regions. The tendency for regions with $\xi=O(1)$ to most frequently coincide with regions where $\tilde{B}_z/N<1$ becomes stronger as the scale is increased. The most important implication of this is that the dynamical significance of the alignments on the flow energetics the we have demonstrated cannot be understood through a simple connection between the local alignments and local stability of the flow.

One important aspect for future work is to consider how the transport terms in the TKE and APE equations depend on $\xi$. Even though these terms have unconditional averages that are zero for a statistically homogeneous flow, their averages conditioned on $\xi$ need not be zero. Analyzing such conditional averages would provide important insights into the local role of the transport mechanisms and how their role is influenced by the ramp-cliff structures in the density field. The results from the present work also suggest that models for stratified turbulence that explicitly resolve alignments between various gradient fields would be valuable for predicting the spatio-temporal variability of mixing rates in stratified turbulence. In particular, this would be valuable since regions where mixing rates are much larger than the mean value were shown to occur preferentially in regions where $\xi\approx -1$, especially for $Pr=1$. One possible avenue would be models based on the Lagrangian evolution of velocity and density gradients in turbulent flows, such as have already been developed for the neutrally buoyant regime \citep{Zhang2023}.

\backsection[Acknowledgements]
{This research used resources of the Oak Ridge Leadership Computing Facility at the Oak Ridge National Laboratory, which is supported by the Office of Science of the U.S. Department of Energy under Contract No. DE-AC05-00OR22725.  Additional resources were provided
through the U.S.\ Department of Defense High Performance Computing Modernization
Program by the Army Engineer Research and Development Center and the Army
Research Laboratory under Frontier Project FP-CFD-FY14-007.
S.B. is grateful to Miles Couchman and Young Ro Yi for their insightful feedback.}
\backsection[Funding]
{S.B. and A.D.B. were supported by National Science Foundation (NSF) CAREER award \# 2042346. }

\backsection[Declaration of interests]
{The authors report no conflict of interest.}

\backsection[Author ORCIDs]{Soumak Bhattacharjee, https://orcid.org/0000-0002-3123-8973; Stephen M. de Bruyn Kops https://orcid.org/0000-0002-7727-8786; Andrew D. Bragg, https://orcid.org/0000-0001-7068-8048}

\bibliographystyle{jfm}
\bibliography{jfm,Paper3}

\newpage


\end{document}